\RequirePackage{fix-cm}
\documentclass[smallcondensed]{svjour3}     
\smartqed  

\usepackage{mathptmx}      
\usepackage{graphicx}
\usepackage{amsmath}
\usepackage{amssymb}
\usepackage{setspace}
\usepackage{enumerate}
\usepackage[ruled,linesnumbered,vlined,algo2e,resetcount]{algorithm2e}
\graphicspath{{figs/}}
\usepackage{color}
\usepackage{array}
\newcolumntype{C}[1]{>{\centering\let\newline\\\arraybackslash\hspace{0pt}}m{#1}}
\usepackage{microtype}
\usepackage{hyperref}
\usepackage{cite}
\usepackage{multirow}

\newcounter{exctr}

\journalname{ }

\begin{document}

\title{Free-boundary conformal parameterization of point clouds}


\author{Gary P. T. Choi \and Yechen Liu \and Lok Ming Lui}

\institute{
Gary P. T. Choi \at
Department of Mathematics, Massachusetts Institute of Technology, Cambridge, MA, USA \\
\email{ptchoi@mit.edu}           \\
\href{https://orcid.org/0000-0001-5407-9111}{\includegraphics[scale=0.45]{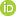}} \href{https://orcid.org/0000-0001-5407-9111}{https://orcid.org/0000-0001-5407-9111}
\and
Yechen Liu \at
Department of Applied Mathematics and Statistics, Johns Hopkins University, Baltimore, MD, USA\\
\email{yliu381@jhu.edu}
\and
Lok Ming Lui \at
Department of Mathematics, The Chinese University of Hong Kong, Kong Kong\\
\email{lmlui@math.cuhk.edu.hk}           \\
\href{https://orcid.org/0000-0002-9152-0743}{\includegraphics[scale=0.45]{orcid}} \href{https://orcid.org/0000-0002-9152-0743}{https://orcid.org/0000-0002-9152-0743}
}

\date{Received: date / Accepted: date}

\maketitle
\begin{abstract}
With the advancement in 3D scanning technology, there has been a surge of interest in the use of point clouds in science and engineering. To facilitate the computations and analyses of point clouds, prior works have considered parameterizing them onto some simple planar domains with a fixed boundary shape such as a unit circle or a rectangle. However, the geometry of the fixed shape may lead to some undesirable distortion in the parameterization. It is therefore more natural to consider free-boundary conformal parameterizations of point clouds, which minimize the local geometric distortion of the mapping without constraining the overall shape. In this work, we develop a free-boundary conformal parameterization method for disk-type point clouds, which involves a novel approximation scheme of the point cloud Laplacian with accumulated cotangent weights together with a special treatment at the boundary points. With the aid of the free-boundary conformal parameterization, high-quality point cloud meshing can be easily achieved. Furthermore, we show that using the idea of conformal welding in complex analysis, the point cloud conformal parameterization can be computed in a divide-and-conquer manner. Experimental results are presented to demonstrate the effectiveness of the proposed method.

\keywords{conformal parameterization \and point cloud \and free-boundary \and Laplace--Beltrami operator \and conformal welding}

\subclass{65D18 \and 68U05 \and 52C26 \and 68W10}
\end{abstract}

\section{Introduction}
With the rapid development of computer technology, the acquisition and use of geometric data have become increasingly popular~\cite{rusu20113d}. The simplest form of geometric data obtained by 3D scanners is a set of points in $\mathbb{R}^3$, which is also known as a \emph{point cloud}. Point clouds have been widely studied for 3D modeling~\cite{remondino2003point,mitra2004registration}, object detection~\cite{schnabel2007efficient}, shape analysis~\cite{collins2004barcode} etc. and have recently become a subject of interest in machine learning~\cite{shen2018mining,zhou2018voxelnet,shi2019pointrcnn,liu2019relation}. However, working with point clouds in the three-dimensional space is usually complicated and computationally expensive. It is therefore desirable to have a method for projecting the point clouds onto a lower dimensional space without distorting their shape, such that the computations can be further simplified.

Surface parameterization is the process of mapping a complicated surface onto a simpler domain. Over the past several decades, numerous efforts have been devoted to the development of surface parameterization algorithms with applications in science and engineering. In general, any parameterization must unavoidably induce certain distortion in area, angle, or both. Therefore, two major classes of surface parameterization methods are area-preserving (authalic) parameterizations and angle-preserving (conformal) parameterizations. Prior area-preserving parameterization methods include Lie advection~\cite{zou2011authalic}, optimal mass transport (OMT)~\cite{zhao2013area,su2016area,pumarola20193dpeople,giri2020open}, density-equalizing map (DEM)~\cite{choi2018density,choi2020area}, stretch energy minimization (SEM)~\cite{yueh2019novel} etc. These methods focus on preserving the size of the area elements but not their shape. Previous works on conformal parameterization include harmonic energy minimization~\cite{pinkall1993computing,gu2004genus}, least-square conformal map (LSCM)~\cite{levy2002least}, discrete natural conformal parameterization (DNCP)~\cite{desbrun2002intrinsic}, angle-based flattening (ABF)~\cite{sheffer2001parameterization,sheffer2005abf}, Yamabe flow~\cite{luo2004combinatorial}, circle patterns~\cite{kharevych2006discrete}, spectral conformal map~\cite{mullen2008spectral}, Zipper algorithm~\cite{marshall2007convergence}, Ricci flow~\cite{jin2008discrete,yang2009generalized}, boundary first flattening~\cite{sawhney2017boundary}, conformal energy minimization~\cite{yueh2017efficient} etc. (see~\cite{floater2005surface,sheffer2006mesh,hormann2007mesh} for detailed surveys on the subject). These parameterization methods preserve angles and hence the local geometry of the surfaces, which is desirable in many applications. Many of them (e.g.~\cite{levy2002least,mullen2008spectral,sawhney2017boundary}) also allow the boundary of the parameter domain to vary from a standard shape and achieve a more flexible parameterization result. In recent years, quasi-conformal theory has been utilized for conformal parameterization~\cite{choi2015flash,choi2015fast,choi2017conformal,choi2018linear,choi2020parallelizable,choi2020efficient} and applied to surface remeshing~\cite{choi2016fast,choi2017subdivision}, image registration~\cite{lui2014teichmuller,yung2018efficient}, biological shape analysis~\cite{choi2018planar,choi2020tooth,choi2020shape} and material design~\cite{choi2019programming}. However, most of the above parameterization methods only work for surface meshes with the structural connectivity prescribed. 

\begin{figure}[t]
    \centering
    \includegraphics[width=0.95\textwidth]{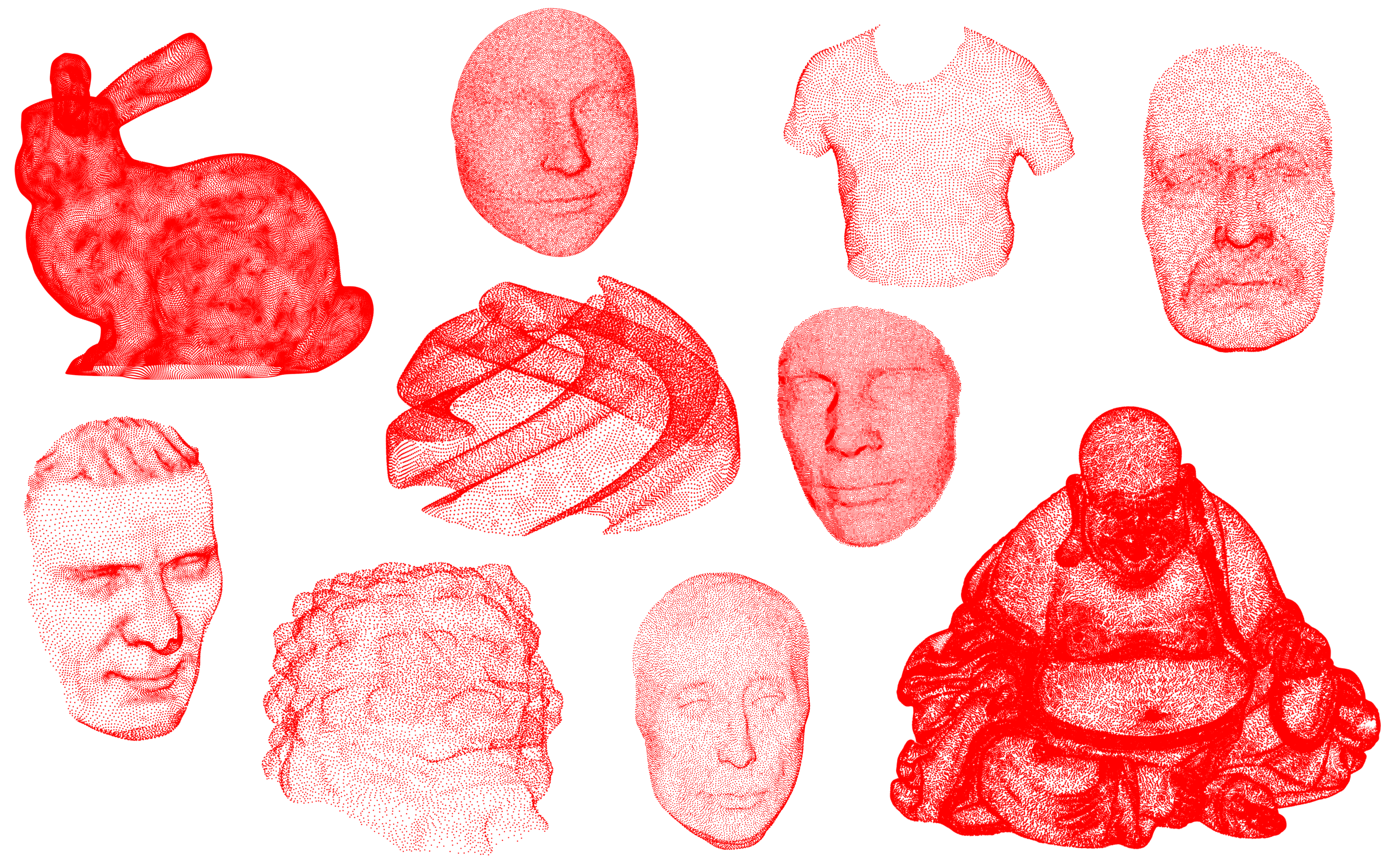}
    \caption{A gallery of disk-type point clouds. As the underlying surface of each of them is a simply-connected open surface, it is natural to consider the conformal parameterization of these point clouds onto planar domains with a single boundary. Moreover, the large variation of the point cloud boundary shapes suggests the need of a free-boundary conformal parameterization method which takes the boundary shapes into consideration.}
    \label{fig:pc_gallery}
\end{figure}

Unlike surface meshes, point clouds do not contain any information of the connectivity of the points and hence are more difficult to handle in general. There have only been a few works on the parameterization of point cloud data~\cite{zwicker2004meshing,tewari2006meshing,zhang2010mesh,meng2013parameterization,choi2016spherical,meng2016tempo,sharp2020laplacian}. In particular, to compute the conformal parameterization of a point cloud, it is common to approximate the Laplace--Beltrami operator at every vertex using integral approximation~\cite{belkin2008towards,belkin2009constructing}, the moving least-square (MLS) method~\cite{wendland1995piecewise,liang2012geometric,liang2013solving} or the local mesh method~\cite{lai2013local}, and then solve the Laplace equation with some boundary constraints. However, most of the existing approximation schemes only work well for the case of fixed boundary constraints, in which the boundary shape of the target parameter domain is usually set to be either a circle or a rectangle. Enforcing such a fixed boundary shape creates undesirable geometric distortion in the parameterization result. A possible remedy is to consider a \emph{free-boundary} conformal parameterization, in which the positions of only two boundary points are fixed for eliminating translation, rotation and scaling, and each of all the remaining boundary points is automatically mapped to a suitable location according to the geometry of the given point cloud. In this work, we develop a free-boundary conformal parameterization method for disk-type point clouds (Fig.~\ref{fig:pc_gallery}). In particular, we construct the point cloud Laplacian by accumulating cotangent weights at different local Delaunay triangulations. Also, as the free-boundary parameterization relies heavily on the approximation of the Laplacian at the boundary, we propose a new approximation scheme with a novel angle criterion for handling the approximation at the boundary points. The parameterization method can be utilized for high-quality point cloud meshing. Furthermore, by extending the idea of partial welding~\cite{choi2020parallelizable}, we can solve the parameterization problem by decomposing the point cloud into subdomains, solving the parameterization for each of them, and finally gluing them seamlessly. This approach can largely simplify the computation for parameterizing dense point clouds while preserving the conformality of the mapping.

The rest of the paper is organized as follows. In Section~\ref{sect:contributions}, we highlight the contributions of our work. In Section~\ref{sect:background}, we review some mathematical concepts related to our work. The proposed method is then described in detail in Section~\ref{sect:main}. Experimental results are presented in Section~\ref{sect:experiment} to demonstrate the effectiveness of our proposed method. In Section~\ref{sect:application}, we describe the application of our method to point cloud meshing. In Section~\ref{sect:extension}, we discuss the extension of our proposed method using the idea of partial welding. We conclude the paper and discuss possible future works in Section~\ref{sect:discussion}. 

\section{Contributions} \label{sect:contributions}
The contributions of our work are three-fold:
\begin{enumerate}[(i)]
\item  We develop a free-boundary conformal parameterization method for disk-type point clouds. Our method involves a new approximation scheme for the point cloud Laplacian with a novel angle criterion for handling the possible non-convexity of the point cloud boundary. Experimental results show that the proposed angle criterion leads to a significant improvement in the conformality of the parameterization.
 \item Using the proposed free-boundary parameterization method, we can easily produce high-quality triangular meshes for disk-type point clouds. 
 \item We further extend the proposed parameterization method using the idea of partial welding, which enhances the flexibility of the computation of the free-boundary conformal parameterization of point clouds.
\end{enumerate}

\section{Mathematical background} \label{sect:background}
\subsection{Harmonic maps}
Let $S$ be a surface in $\mathbb{R}^3$. A map $f: S\rightarrow\mathbb{R}^2$ is said to be a \emph{harmonic map} if it is a critical point of the following Dirichlet energy~\cite{pinkall1993computing}:
\begin{equation}
    E_D(f) = \dfrac{1}{2}\int_S |\nabla f|^2dA.
    \label{eqt:dirichletenergy}
\end{equation}
$f$ is also the solution to the Laplace equation
\begin{equation}
    \Delta f = 0.
\end{equation}
To see this, one can consider the following energy
\begin{equation}
    E(f,v) = \int_S \nabla f\cdot \nabla v.
\end{equation}
Let $u$ be a critical point to Eq.~\eqref{eqt:dirichletenergy} and $v$ be a test function that vanishes on $\partial S$. By differentiating $E(u,u+tv)$ with respect to $t$, one can show that $u$ minimizes $E_D$ by the Stoke's theorem or integration by part.

\subsection{Conformal maps}
Let $U\subset\mathbb{C}$. A map $f: U\rightarrow\mathbb{C}$ is said to be a \emph{conformal map} if $f(x,y) = u(x,y)+iv(x,y)$ satisfies the Cauchy--Riemann equations:
\begin{equation}
\left\{\begin{array}{ll}
    \dfrac{\partial u}{\partial x} &= \dfrac{\partial v}{\partial y},\\
    \dfrac{\partial u}{\partial y} &= -\dfrac{\partial v}{\partial x}.
\end{array} \right.
    \label{eqt:CauchyRiemann}
\end{equation}
Consequently, a conformal map can be viewed as a critical point of the following energy:
\begin{equation}
    E_C(f)=\dfrac{1}{2}\int_U\left[\left(\dfrac{\partial u}{\partial x}-\dfrac{\partial v}{\partial y}\right)^2 + \left(\dfrac{\partial u}{\partial y}+\dfrac{\partial v}{\partial x}\right)^2\right]dA.
    \label{eqt:conformalenergy}
\end{equation}

Note that conformal maps preserve angles and hence the local geometry of the shape, with infinitesimal circles mapped to infinitesimal circles. To see this, let $\gamma_i: [-\epsilon,\epsilon]\rightarrow U, \epsilon>0$ with $i=1,2$ be two curves satisfying that $\gamma_i(0) = z$ with $\gamma_i'(0) = v_i, i=1,2$. We have
\begin{equation}
\dfrac{(f\circ\gamma_1)'(0)(f\circ\gamma_2)'(0)}{\|(f\circ\gamma_1)'(0)(f\circ\gamma_2)'(0)\|} = \dfrac{[f'(z)]^2\gamma_1'(0)\gamma_2'(0)}{\|f'(z)\|^2\|\gamma_1'(0)\gamma_2'(0)\|} = \dfrac{v_1v_2}{\|v_1v_2\|}.
\end{equation} 
This shows the angle-preserving property of conformal maps.

Conformal maps and harmonic maps are closely related. Note that if $S$ is a simply-connected open surface, it can be represented using a single chart $(U, \phi)$. Then, the concept of conformal maps can be naturally extended for surfaces. Now, by rewriting Eq.~\eqref{eqt:dirichletenergy} in the following form
\begin{equation}
    E_D(f) = \dfrac{1}{2}\int_U |\nabla f|^2dA = \dfrac{1}{2}\int_U (u_x^2+u_y^2+v_x^2+v_y^2)dA,
\end{equation}
one can see that
\begin{equation}
    E_C(f)-E_D(f) = \int_U\left(u_yv_x-u_xv_y\right)dA = -\int_U \left(\dfrac{\partial f}{\partial x}\times \dfrac{\partial f}{\partial y} \right)dA.
\end{equation}
Note that $\left(\dfrac{\partial f}{\partial x}\times \dfrac{\partial f}{\partial y} \right)dA$ is the area element of $f(U)$. Hence, the right-hand side of the above equation is the total area of $f(U)$. Denoting it by $A(f)$, we have 
\begin{equation}
    E_C(f) = E_D(f) - A(f) \geq 0.
    \label{eqt:relation}
\end{equation}
Since $f$ is conformal if and only if $E_C(f)=0$, conformal maps can be viewed as the harmonic maps achieving the maximum area. 

\subsection{M\"obius transformation}
A \emph{M\"obius transformation} $f:\mathbb{C}\rightarrow\mathbb{C}$ is a conformal map on the (extended) complex plane in the form
\begin{equation}
    f(z) = \dfrac{az+b}{cz+d}, 
\end{equation}{}
with $a,b,c,d\in\mathbb{C}$ satisfying $ad-bc\neq 0$. 

It can be observed that $f$ maps the three points $(0, -\frac{d}{c}, \infty)$ to $(\frac{b}{d}, \infty, \frac{a}{c})$ on the extended complex plane. By making use of the three-point correspondence, one can utilize M\"obius transformations for transforming a planar shape into some desired target shape while preserving conformality.

\section{Proposed method} \label{sect:main}
Given a point cloud $\mathcal{P}$ representing a simply-connected open surface, our goal is to find a free-boundary conformal parameterization $f: \mathcal{P} \to \mathbb{R}^2$. Below, we first introduce a free-boundary conformal parameterization method for triangulated surfaces. We then extend it for parameterizing point cloud surfaces in a free-boundary manner by proposing a new approximation scheme for the point cloud Laplacian. In particular, a novel angle criterion is used for improving the approximation at the point cloud boundary and yielding a more accurate conformal parameterization result.

\subsection{Discrete natural conformal parameterization (DNCP) for triangulated surfaces}
The discrete natural conformal parameterization (DNCP) method~\cite{desbrun2002intrinsic} is a method for computing free-boundary conformal parameterizations of triangulated surfaces. Let $S = (\mathcal{V}, \mathcal{E}, \mathcal{F})$ be a simply-connected open triangulated surface, where $\mathcal{V}= \{v_i\}_{1\leq i\leq n}$ is the set of vertices, $\mathcal{E}$ is the set of edges, and $\mathcal{F}$ is the triangulation. The DNCP method finds the desired conformal parameterization $f:S \to \mathbb{R}^2$ by linearizing Eq.~\eqref{eqt:relation}. More specifically, the Dirichlet energy $E_D(f)$ in Eq.~\eqref{eqt:dirichletenergy} can be discretized using the cotangent formula~\cite{pinkall1993computing}:
\begin{equation}
    E_D(f) = \dfrac{1}{2}\sum\limits_{(p,q): [v_p, v_q] \in \mathcal{E}} \frac{\cot{\alpha_{pq}}+\cot{\beta_{pq}}}{2}|f(v_p)-f(v_q)|^2,
\end{equation}
where $\alpha_{pq},\beta_{pq}$ are the opposite angles of the edge $[v_p,v_q]$. If we write $f$ as a $2n \times 1$ vector with $f =  (f_x, f_y)^T = (x_1, x_2, \dots x_n, y_1, y_2,\dots,y_n)^T$ where $f(v_i)=(x_i, y_i)^T$, then we have
\begin{equation}
E_D(f) = \dfrac{1}{2} f^T \left(
\begin{array}{c|c}
L & 0 \\
\hline
0 & L
\end{array}
\right) f,
\label{eqt:Cotangentmatrix}
\end{equation}
where $L$ is an $n\times n$ matrix with
\begin{equation}
L_{ij} = 
\begin{cases}
-\dfrac{1}{2}(\cot{\alpha_{ij}}+\cot{\beta_{ij}}), \hspace{12pt} &\text{ if } [v_i,v_j] \in \mathcal{E}, \\
 -\sum\limits_{m\neq i} L_{im} = \dfrac{1}{2}\sum\limits_{m: [v_i, v_m] \in \mathcal{E}}(\cot{\alpha_{im}}+\cot{\beta_{im}})  &\text{ if } i=j,\\
0 \hspace{24pt}&\text{ otherwise}. \\
\end{cases}
\label{eqt:Cotangentweights}
\end{equation}

The area term $A(f)$ in Eq.~\eqref{eqt:relation} can be discretized by considering all edges on the surface boundary $\partial S$:
\begin{equation}
    A(f) = \dfrac{1}{2}\sum\limits_{(v_i,v_j) \in \partial S} (x_iy_j-x_jy_i) = \frac{1}{2}(x^T y^T)M^{area}\begin{pmatrix}
      x  \\
        y 
    \end{pmatrix}.
    \label{eqt:area_term}
\end{equation}
Here, $M^{area}$ is a $2n \times 2n$ matrix in the form of
\begin{equation}
    M^{area} = \left(\begin{array}{c|c}
    0 & M_1 \\
    \hline
    M_2 & 0
    \end{array}
    \right),
\label{eqt:areamatrix}
\end{equation}
where $M_1(i,j) = M_2(j,i) = \frac{1}{2}$ and $M_1(j,i)=  M_2(i,j) = -\frac{1}{2}$ if $(v_i,v_j)$ is an edge on the boundary with positive orientation. 

Altogether, Eq.~\eqref{eqt:relation} can be discretized and rewritten in the following matrix form:
\begin{equation}
    E_C(f) = \dfrac{1}{2} (f_x^T f_y^T)\left(\left(
\begin{array}{c|c}
L & 0 \\
\hline
0 & L
\end{array}
\right)-M^{area} \right)\begin{pmatrix}
      f_x \\ f_y \end{pmatrix}.
\end{equation}
Minimizing $E_C(f)$ is then equivalent to solving the following matrix equation:
\begin{equation}
\Bigg(\left(
\begin{array}{c|c}
L & 0 \\
\hline
0 & L
\end{array}
\right) - M^{area}\Bigg) \begin{pmatrix}
      f_x \\ f_y \end{pmatrix} = 0.
     \label{eqt:DNCP}
\end{equation}
To remove the freedom of rigid motions and scaling, DNCP adds two boundary constraints to map the farthest two vertices in $\mathcal{V}$ to $(0,0)$ and $(1,0)\subset\mathbb{R}^2$. Readers are referred to~\cite{desbrun2002intrinsic} for more details.

As a remark, in general the Laplace--Beltrami operator for triangulated surfaces is discretized as $\Delta = M^{-1} L$, where $L$ is the cotangent Laplacian and $M$ is a mass matrix for normalizing area, which is usually constructed based on Voronoi or barycentric cells (see~\cite{reuter2009discrete} for more details). However, in the above-mentioned formulation of the free-boundary parameterization problem, $\Delta$ is approximated using the cotangent Laplacian only (i.e. with $M$ being an identity matrix). In other words, every vertex is considered to be with unit mass. In fact, this is consistent with the discretization of the area term $A(f)$ in Eq.~\eqref{eqt:area_term}, in which the boundary vertices are also treated to be with uniform weight. If we use a Voronoi or barycentric mass matrix $M$ for the Laplace--Beltrami discretization, the matrices $M_1, M_2$ in the area matrix $M^{area}$ in Eq.~\eqref{eqt:areamatrix} should also be replaced with $M^{-1} M_1$ and $M^{-1}M_2$. Then, Eq.~\eqref{eqt:DNCP} becomes 
$
\left(
\begin{array}{c|c}
M^{-1} L & -M^{-1} M_1 \\
\hline
-M^{-1} M_2 & M^{-1} L
\end{array}
\right) \begin{pmatrix}
      f_x \\ f_y \end{pmatrix} = 0,
$
from which it is clear that the choice of the mass matrix $M$ does not affect the solution. Therefore, one can simply use the cotangent Laplacian for the Laplace--Beltrami discretization in the DNCP method.

\subsection{Point cloud Laplacian with angle-based convexity modification}
For our problem of free-boundary conformal parameterization of point clouds, the above mesh-based discretization cannot be directly applied. In particular, the cotangent Laplacian in Eq.~\eqref{eqt:Cotangentmatrix} cannot be constructed because of the absence of the connectivity information in point clouds. To resolve this issue, a possible way is to develop an alternative approximation of the Laplacian. Note that the cotangent Laplacian only involves the neighbors of every vertex. This motivates us to consider reconstructing the local geometric structure at every vertex of the point cloud and approximating the Laplacian using the local structure. More specifically, we construct the point cloud Laplacian by accumulating cotangent weights obtained from different local Delaunay triangulations. Moreover, as the free-boundary conformal parameterization relies heavily on the approximation of the point cloud Laplacian at the boundary, we further develop a novel angle-based convexity modification scheme for handling the Laplacian approximation at the boundary points. Altogether, this allows us to achieve an accurate point cloud parameterization result.

To obtain the accumulated cotangent weights, we make use of the k-nearest-neighbors (kNN) algorithm, as well as the principal component analysis (PCA) method and the Delaunay triangulation method. Let $\mathcal{P} = \{v_i\}_{i=1}^n$ be a point cloud surface with an oriented boundary $\Gamma = \partial \mathcal{P}$, and $k$ be a prescribed kNN parameter. We first find the $k$-nearest neighbors $N^k_i = \{v_{n_1},v_{n_2},\dots,v_{n_k}\}$ of each vertex $v_i$. To capture the local geometric information around $v_i$, we apply PCA to find the three principal directions $\{e_i^1,e_i^2,e_i^3\}$ of these $k$ data points, and take the plane $\text{span}(e_i^1,e_i^2)$ passing through $v_i$ as the tangent plane of $v_i$. We project $N^k_i$ to the tangent plane and get $\widetilde{N}^k_i = \{\widetilde{v}_{n_j}\}_{j=1}^k$, using the projection formula 
\begin{equation} \label{eqt:projection}
    \widetilde{v}_{n_j} = v_{n_j} - \langle v_{n_j}, e_i^3 \rangle e_i^3.
\end{equation}
Then, we construct the 2D Delaunay triangulation for $\widetilde{N}^k_i$. Note that the Delaunay triangulation maximizes the minimal angle of each triangle and the construction algorithm is provably convergent. Consequently, we can obtain a nice triangulation representing the local geometric structure around each vertex. Based on the local Delaunay triangulation around each $v_i$, we obtain the one-ring neighborhood $R_i$ of $v_i$. We can then apply the cotangent formula in Eq.~\eqref{eqt:Cotangentweights} to construct an $n\times n$ matrix $L^{pc}_{k,i}$ using the angles in $R_i$. More explicitly, we have
\begin{equation}
\begin{cases}
L^{pc}_{k,i}(i,j) = L^{pc}_{k,i}(j,i) = -\dfrac{1}{2}(\cot{\alpha_{ij}}+\cot{\beta_{ij}}) \ \ \text{ if } v_j \in R_i,\\
L^{pc}_{k,i}(i,i) = \dfrac{1}{2}\sum\limits_{j: v_j \in R_i} (\cot{\alpha_{ij}}+\cot{\beta_{ij}}),
\end{cases}
\label{eqt:Cotangentweights_pc}
\end{equation}
and all other entries of $L^{pc}_{k,i}$ are set to be zero. Unlike the matrix $L$ in Eq.~\eqref{eqt:Cotangentweights}, which is computed based on the entire triangulated surface, our matrix $L^{pc}_{k,i}$ only covers the local structure around $v_i$. Therefore, the point cloud Laplacian for $\mathcal{P}$ can be approximated by accumulating the cotangent weights in all $L^{pc}_{k,i}$, with $i = 1, 2, \dots, n$.

\begin{figure}[t]
    \centering
    \includegraphics[width=\textwidth]{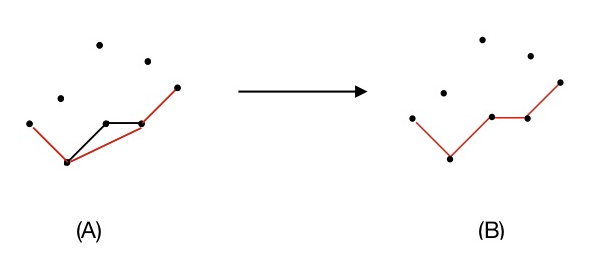}
    \caption{Capturing the concavity of the point cloud boundary. (A) Boundary points captured without applying the angle criterion. (B) Boundary points captured with the angle criterion applied.}
    \label{fig:bd_concavity}
\end{figure}

While the above local Delaunay-based method gives a good approximation of the point cloud Laplacian at the interior vertices, the approximation at the point cloud boundary $\partial \mathcal{P}$ may be inaccurate due to the concavity of boundary points. For instance, if $\partial \mathcal{P}$ contains a concave corner at a vertex $v_i$, the one-ring neighborhood at $v_i$ will likely create a convex boundary by wrongly connecting some of its neighboring boundary points under the above-mentioned approximation process (see Fig.~\ref{fig:bd_concavity}(A)), thereby causing inaccuracy in $L^{pc}_{k,i}$. For fixed-boundary parameterization problems, such inaccuracies may not affect the parameterization result as there will be fixed-boundary constraints for all boundary points. However, for our free-boundary conformal parameterization problem, there are only two fixed boundary points in computing the parameterization and hence the result can be significantly affected by the inaccurately approximated Laplacian at the boundary. This motivates us to develop a novel scheme for handling the approximation at the point cloud boundary.

More specifically, note that the above-mentioned issue is due to the possible non-convexity of the point cloud boundary. At such non-convex regions, the Delaunay triangulation will produces certain sharp triangles. To correct this, we remove those triangles that contain an angle which is either too large or too small (see Fig.~\ref{fig:bd_concavity}(B)). This can be done by prescribing an angle range $(c_1,c_2)$ and checking if every boundary angle $\theta$ satisfies the following angle criterion:
\begin{equation}
    c_1 <\theta < c_2.
\end{equation}
After removing all those triangles that violate this angle criterion, we obtain the matrices $L^{pc}_{k,i}$ for all the boundary points.

With all $n$ matrices $L^{pc}_{k,1}, L^{pc}_{k,2}, \dots, L^{pc}_{k,n}$ computed, we assemble them to form the approximation of the point cloud Laplacian for the entire $\mathcal{P}$. It is noteworthy that a triangulation is constructed locally at each vertex, and so the triangulations at all points together contain overlapping triangles. As each triangle has three vertices, most of the angles are considered three times in the collection of all $L^{pc}_{k,i}$. For instance, 
suppose $[v_p, v_q, v_r]$ is a triangle in $\mathcal{N}^{\mathcal{F}}(v_p)$, i.e. the 1-ring Delaunay triangulation around the vertex $v_p$. In the Laplacian matrix $L_{k,p}^{pc}$, this triangle contributes a weight to the 9 entries $(p,p)$, $(p,q)$, $(p,r)$, $(q,p)$, $(q,q)$, $(q,r)$, $(r,p)$, $(r,q)$, $(r,r)$. Since having a triangle $[v_p, v_q, v_r]$ in $\mathcal{N}^{\mathcal{F}}(v_p)$ implies that $v_p, v_r$ are close to $v_q$, it is likely that the points $v_p, v_r$ are in the 1-ring vertex neighborhood $\mathcal{N}^{\mathcal{V}}(v_q)$ of the vertex $v_q$. Similarly, it is likely that $v_p, v_q \in \mathcal{N}^{\mathcal{V}}(v_r)$. Hence, it is likely that $[v_p, v_q, v_r] \in ^{\mathcal{F}}(v_q)$ and $[v_p, v_q, v_r] \in ^{\mathcal{F}}(v_r)$ and so it will contribute a weight to the same 9 entries in each of $L_{k,q}^{pc}$ and $L_{k,r}^{pc}$. In other words, the same cotangent weights are likely counted three times. Hence, we obtain the approximated point cloud Laplacian $L^{pc}_k$ by summing up all $L^{pc}_{k,i}$ and dividing it by 3:
\begin{equation} \label{eqt:LBoperator_pc}
    L^{pc}_k = \frac{1}{3}\sum_{i=1}^n L^{pc}_{k,i}.
\end{equation}

We remark that there are two major differences between our approximation and the prior approximation methods. First, a difference between our method and the moving least-square (MLS) method~\cite{liang2012geometric} or the local mesh method~\cite{lai2013local} is in the consideration of \emph{overlapping triangulations}. The MLS method approximates the derivatives locally at each vertex $v_i$ by fitting a local patch of $v_i$ using a combination of polynomials with some prescribed weight functions, in which no triangulations are considered. While the local mesh method also approximates the Laplace--Beltrami operator by constructing a local triangulation at each vertex $v_i$ and considering its one-ring neighborhood $R_i$, the triangulation at $v_i$ only affects the values at the $i$-th row of the Laplacian matrix it creates: 
\begin{equation}
    \Delta(f(v_i)) = \sum_{j: v_j \in R_i} w_{ij} (f(v_j) - f(v_i)),
\end{equation}
where $w_{ij}$ is the cotangent weight. In other words, the approximations at two neighboring points $v_i, v_j$ are handled separately in the local mesh method without any coupling procedure. By contrast, in our proposed approximation scheme, the matrix $L^{pc}_{k,i}$ obtained from the local triangulation at $v_i$ contains nonzero entries not only at the $i$-th row but also at $L^{pc}_{k,i}(j,i)$ for all $j$ with $v_j \in R_i$. The approximations at all points are then coupled by summing up all $L^{pc}_{k,i}$ and dividing it by~3 in Eq.~\eqref{eqt:LBoperator_pc}. Second, when compared to other prior local Delaunay-based point cloud Laplacian approximation schemes with accumulated cotangent weights~\cite{clarenz2004finite,cao2010point,sharp2020laplacian}, our proposed scheme involves an extra step of handling the approximations at boundary vertices using the angle criterion. As we discussed above, this plays an important role for the free-boundary conformal parameterization problem we are considering in this work.

Altogether, with the neighboring geometric information  appropriately coupled at both the interior and boundary of the point cloud and the special treatment for the point cloud boundary, our proposed approximation method yields a better result for the free-boundary conformal parameterization problem. Quantitative comparisons between the approximation schemes are provided in Section~\ref{sect:experiment}.

\subsection{Algorithmic procedure of the proposed point cloud conformal parameterization method}
Using the proposed approximation $L^{pc}_k$ of the point cloud Laplacian constructed in Eq.~\eqref{eqt:LBoperator_pc} and the area matrix $M^{area}$ constructed in Eq.~\eqref{eqt:areamatrix}, we can obtain a free-boundary conformal parameterization $f = (f_x, f_y)^T$ of the point cloud $\mathcal{P}$ by solving a linear system similar to Eq.~\eqref{eqt:DNCP}:
\begin{equation} \label{eqt:DNCP_PC}
    \Bigg(\left(
\begin{array}{c|c}
L^{pc}_k & 0 \\
\hline
0 & L^{pc}_k
\end{array}
\right) - M^{area}\Bigg) \begin{pmatrix}
      f_x \\ f_y \end{pmatrix} = 0.
\end{equation}
For the boundary constraints, we follow the DNCP method~\cite{desbrun2002intrinsic} and map the farthest two points in $\mathcal{P}$ to $(0,0)$ and $(1,0)$. This eliminates the rigid motions and rescaling of the parameterization result and ensures that the overall boundary shape is determined automatically. The proposed method for computing free-boundary conformal parameterizations of point clouds is summarized in Algorithm \ref{algo:DNCP_PC}. 

\begin{algorithm2e}[h!]
    \label{algo:DNCP_PC}
\KwIn{A point cloud $\mathcal{P} = \{v_i\}_{i=1}^n$ with disk topology with (oriented) boundary indices $\partial \mathcal{P} = (b_1,b_2,\dots,b_l)$, the prescribed kNN parameter $k$, and the prescribed angle range $(c_1,c_2)$ (in degrees) for the boundary angles.}
\KwOut{A free-boundary conformal parameterization $f:\mathcal{P} \to \mathbb{R}^2$.}
\BlankLine

\For{$i = 1, \dots, n$}{
    Find the $k$-nearest neighbors $N^k_i =\{v_{n_1},\dots,v_{n_k}\}$ of $v_i$\;
    
    Use PCA to find the first three principal directions $\{e_i^1,e_i^2,e_i^3\}$ of $N^k_i$\;
    
    Project $N^k_i$ to the tangent plane formed by $e_i^1, e_i^2$ passing through $v_i$ and obtain $\widetilde{N}^k_i$ using the projection formula in Eq.~\eqref{eqt:projection}\;
    
    Construct a Delaunay triangulation $T_k^i$ for $\widetilde{N}^k_i$\;
    
    Extract the one-ring neighborhood $R_i$ of $v_i$\;
    
    \If{$v_i\in \partial \mathcal{P} $}{
    Delete those triangles with $\theta\leq c_1$ or $\theta\geq c_2$ and update $R_i$\;
    }
    
    Compute the matrix $L^{pc}_{k,i}$ of the one-ring neighborhoods $R_i$ using cotangent formula in Eq.~\eqref{eqt:Cotangentweights_pc}\;
    }
    
    Obtain the point cloud Laplacian $L^{pc}_k$ using Eq.~\eqref{eqt:LBoperator_pc}\;
    
    Construct the area matrix $M^{area}$ using Eq.~\eqref{eqt:areamatrix}\;
    
    Obtain $f$ by solving the linear system in Eq.~\eqref{eqt:DNCP_PC}\;
    \caption{Free-boundary conformal parameterization of point clouds}
\end{algorithm2e}

\section{Experimental results}\label{sect:experiment}
The proposed algorithm is implemented in MATLAB, with the backslash operator $(\backslash)$ used for solving the linear systems. For the computation of the $k$-nearest neighbors, we use the built-in MATLAB function \texttt{knnsearch}. For the computation of the 2D Delaunay triangulation, we use the built-in MATLAB function \texttt{Delaunay}. The \texttt{parfor} function in the MATLAB parallel computing toolbox is used for speeding up the computation. We adopt point cloud models from online libraries~\cite{stanford,aimatshape} for testing the proposed free-boundary conformal parameterization algorithm. The experiments are performed on a PC with an Intel Core i7-1065G7 quad core CPU and 16~GB RAM. For simplicity, the kNN parameter is set to be $k = 25$ unless otherwise specified. 

\begin{figure}[t!]
    \centering
    \includegraphics[width=\textwidth]{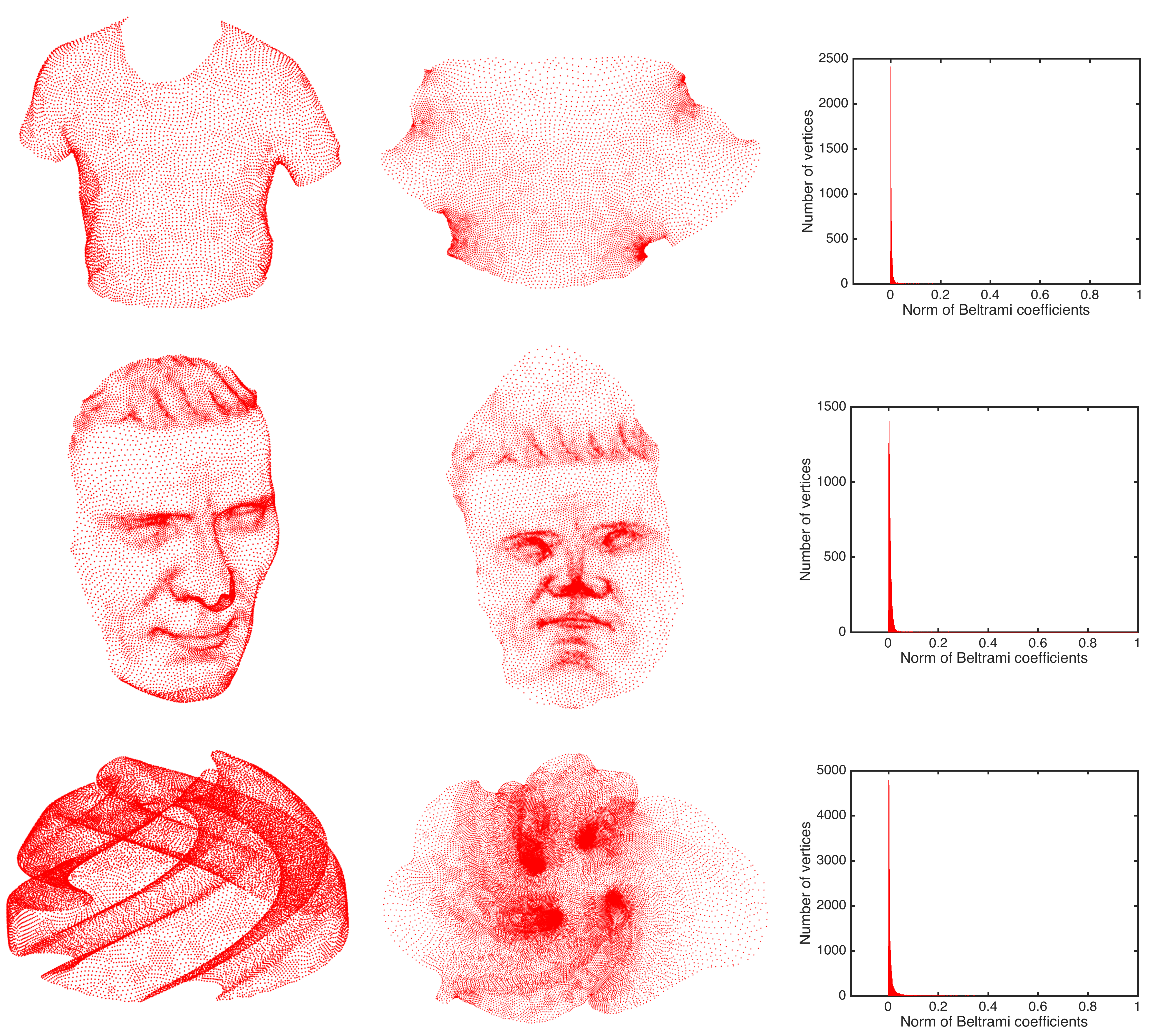}
    \caption{Examples of free-boundary conformal parameterizations of point clouds produced by the proposed method (Algorithm~\ref{algo:DNCP_PC}). Left: The input point clouds. Middle: The parameterization results. Right: The histograms of the norm of the point cloud Beltrami coefficients $|\mu|$.}
    \label{fig:pc_dncp_result}
\end{figure}

\subsection{Free-boundary conformal parameterization of point clouds}
Fig.~\ref{fig:pc_dncp_result} shows three point cloud models and the free-boundary conformal parameterizations achieved using Algorithm~\ref{algo:DNCP_PC}, from which it can be observed that our proposed method is capable of handling point clouds with different geometry. To assess the conformal distortion of each point cloud mapping, we compute the point cloud Beltrami coefficient (PCBC) $\mu$~\cite{meng2018pcbc}, which is a complex-valued function defined on each vertex of the point cloud. In particular, $|\mu| \equiv 0$ if and only if the point cloud mapping is perfectly conformal. As shown in the histograms of $|\mu|$, the parameterizations produced by our proposed method are highly conformal.

\begin{figure}[t!]
    \centering
    \includegraphics[width=0.6\textwidth]{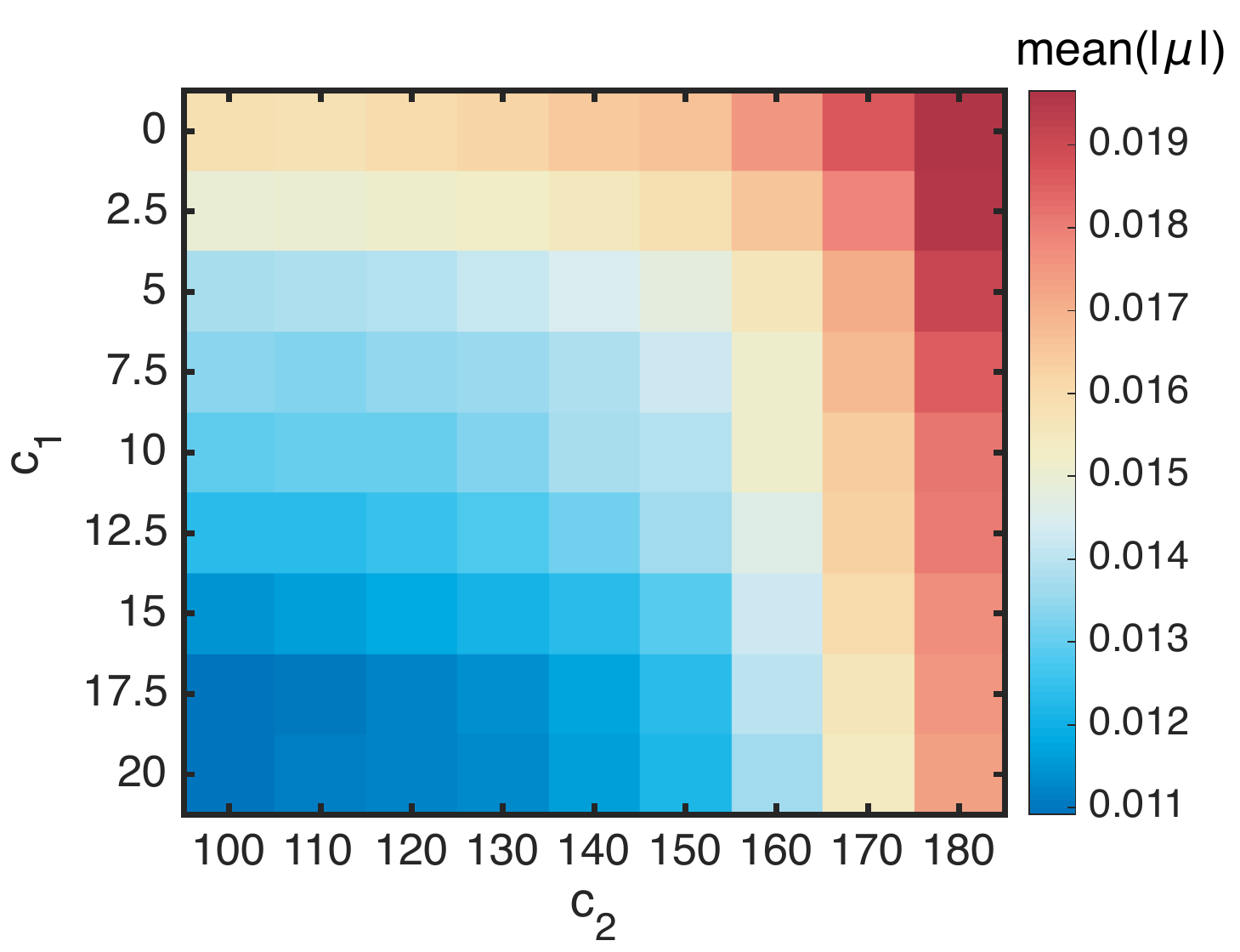}
    \caption{An illustration of the effect of different choices of the angle criterion parameters $(c_1, c_2)$ in Algorithm~\ref{algo:DNCP_PC} on the conformal distortion of the point cloud parameterization. Here, the point cloud used is the first model in Fig.~\ref{fig:pc_dncp_result}.}
    \label{fig:cloth_c1_c2}
\end{figure}

For a more quantitative analysis, Table~\ref{table:free_boundary} records the computation time and the conformal distortion of the proposed method for various point cloud models. For each example, we search for an optimal set of angle criterion parameters $(c_1, c_2)\in [0, 20] \times [100, 180]$ using a simple marching scheme with an increment of $2.5$ for $c_1$ and an increment of $10$ for $c_2$ (see Fig.~\ref{fig:cloth_c1_c2} for an illustration). It can be observed that our method is highly efficient and accurate. For comparison, we also consider running the proposed method without the angle criterion step by setting $(c_1, c_2) = (0, 180)$ (i.e. using the point cloud Laplacian with accumulated cotangent weights~\cite{clarenz2004finite,cao2010point}). The results show that the angle criterion step effectively reduces the conformal distortion by over 30\% on average. This suggests that the proposed angle criterion step is important for yielding an accurate parameterization result.

\begin{figure}[t]
    \centering
    \includegraphics[width=0.9\textwidth]{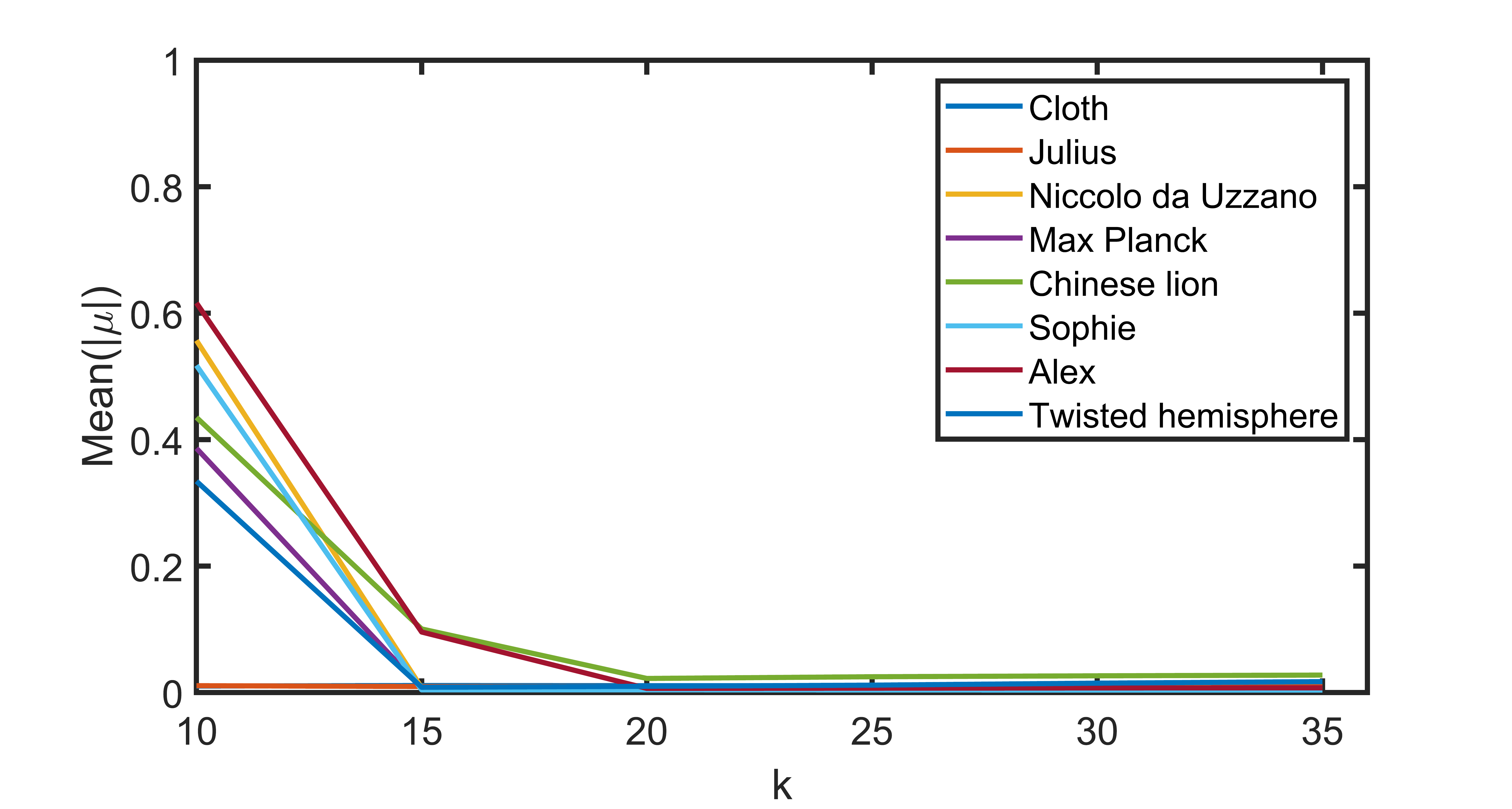}
    \caption{The conformal distortion (in terms of Mean($|\mu|$)) achieved by the proposed free-boundary conformal parameterization method with different choices of the kNN parameter $k$.}
    \label{fig:comparison_k}
\end{figure}

One may be interested in the robustness of the proposed method. Note that the kNN parameter $k$ is used for the construction of the local mesh, in which the 1-ring neighborhood is used for getting the cotangent weights. Using a very small $k$ may lead to inaccuracies in approximating the 1-ring neighborhood, while using a very large $k$ may not be necessary as most of the points will likely be outside the 1-ring neighborhood or even far away from the reference point. In practice, we find that $k=25$ works well for the models we have considered. More specifically, Fig.~\ref{fig:comparison_k} shows the conformal distortion of the parameterization obtained by our proposed method with different $k$. It can be observed that the distortion is usually relatively large when a very small $k$ (e.g. $k=10$) is used. As the value of $k$ increases, the distortion gradually decreases as the local 1-ring neighborhood approximation is more and more accurate. After reaching around $k = 20$, the distortion stabilizes and so $k=25$ is already sufficient for yielding a good parameterization result for all of the models. As for the angles $(c_1, c_2)$ in the angle criterion, note that the experiments presented in Table~\ref{table:free_boundary} cover a large variety of point clouds with different boundary shape, and the optimal values for $(c_1, c_2)$ are similar in all experiments (with $c_1 \approx 15$ and $c_2 \approx 120$). Therefore, we expect that setting similar values for $c_1$ and $c_2$ is sufficient for yielding a notable improvement in conformality for general point clouds. One of the possible future works would be to devise a method for determining the optimal values of $(c_1, c_2)$ directly based on the geometry of the point cloud.

\begin{table}[t]
    \centering
    \begin{tabular}{c|c|c|c|c|C{18mm}|C{20mm}}
        Model & \# points & $c_1$ & $c_2$ & Time (s) & Mean($|\mu|$) (our method) & Mean($|\mu|$) (without the angle criterion) \\ \hline  
        Cloth & 7K & 17.5 & 100 & 1.7 & 0.0109 & 0.0197 \\ 
        Julius & 11K & 20 & 100  & 2.6 & 0.0101 & 0.0155\\ 
        Niccol\`o da Uzzano & 13K & 17.5 & 130 & 3.0 & 0.0079 & 0.0131 \\ 
        Max Planck & 13K  & 20 & 100 & 3.2 & 0.0069 & 0.0110\\ 
        Chinese lion & 17K & 20 & 120 & 5.0 & 0.0251 & 0.0369\\  
        Sophie & 21K & 20 & 120 & 5.4 & 0.0042 & 0.0074\\ 
        Alex & 25K & 7.5 & 100 & 6.7 & 0.0070  & 0.0074\\ 
        Twisted hemisphere & 28K & 12.5 & 120 & 7.8 & 0.0118 & 0.0143 \\ 
    \end{tabular}
    \caption{The performance of the proposed free-boundary conformal parameterization method for point clouds. Here, $\mu$ is the point cloud Beltrami coefficient (PCBC) of the parameterization. For each example, we search for an optimal set of the angle criterion parameters $(c_1, c_2) \in [0, 20] \times [100, 180]$ for Algorithm~\ref{algo:DNCP_PC} using a simple marching scheme. The kNN parameter is fixed to be $k = 25$ for all examples. For comparison, the performance of the parameterization using the local Delaunay-based point cloud Laplacian without the angle criterion (i.e. $(c_1, c_2) = (0, 180)$) is also recorded.}
    \label{table:free_boundary}
\end{table}

\begin{figure}[t!]
    \centering
    \includegraphics[width=0.85\textwidth]{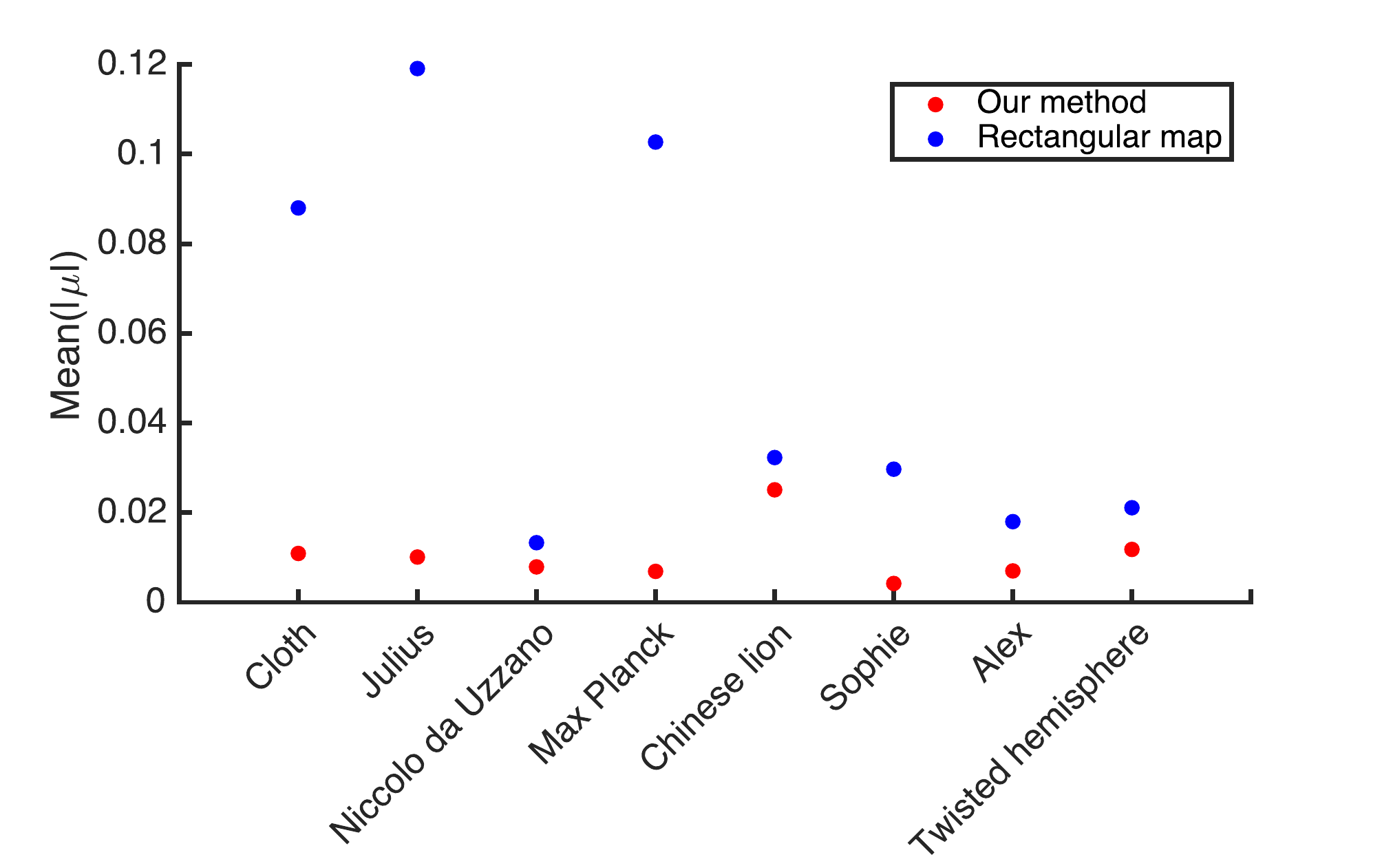}
    \caption{Comparison between our proposed free-boundary conformal parameterization method and the rectangular conformal parameterization method~\cite{meng2016tempo} for point clouds. For each point cloud model, we apply the two methods and record the mean of the norm of the point cloud Beltrami coefficients (PCBC) $|\mu|$.}
    \label{fig:comparison_rect}
\end{figure}

\subsection{Comparison with fixed-boundary conformal parameterization}
After assessing the performance of the proposed free-boundary conformal parameterization method, we compare it with the point cloud rectangular conformal parameterization method~\cite{meng2016tempo}, which maps a disk-type point cloud onto a rectangular domain. As shown in Fig.~\ref{fig:comparison_rect}, our proposed method results in a lower conformal distortion when compared to the rectangular parameterization method. More quantitatively, the distortion by our method is 65\% lower than that by the rectangular parameterization method on average. The better performance of our method can be explained by the fact that while the existence of a conformal map from a simply-connected open surface onto a rectangle is theoretically guaranteed, the additional rectangular boundary constraint may induce numerical inaccuracy in the computation of the point cloud rectangular conformal parameterization. By contrast, the proposed method computes a free-boundary conformal parameterization, in which the input point cloud can be flattened onto the plane more naturally according to their overall geometry.

\begin{figure}[t]
    \centering
    \includegraphics[width=0.75\textwidth]{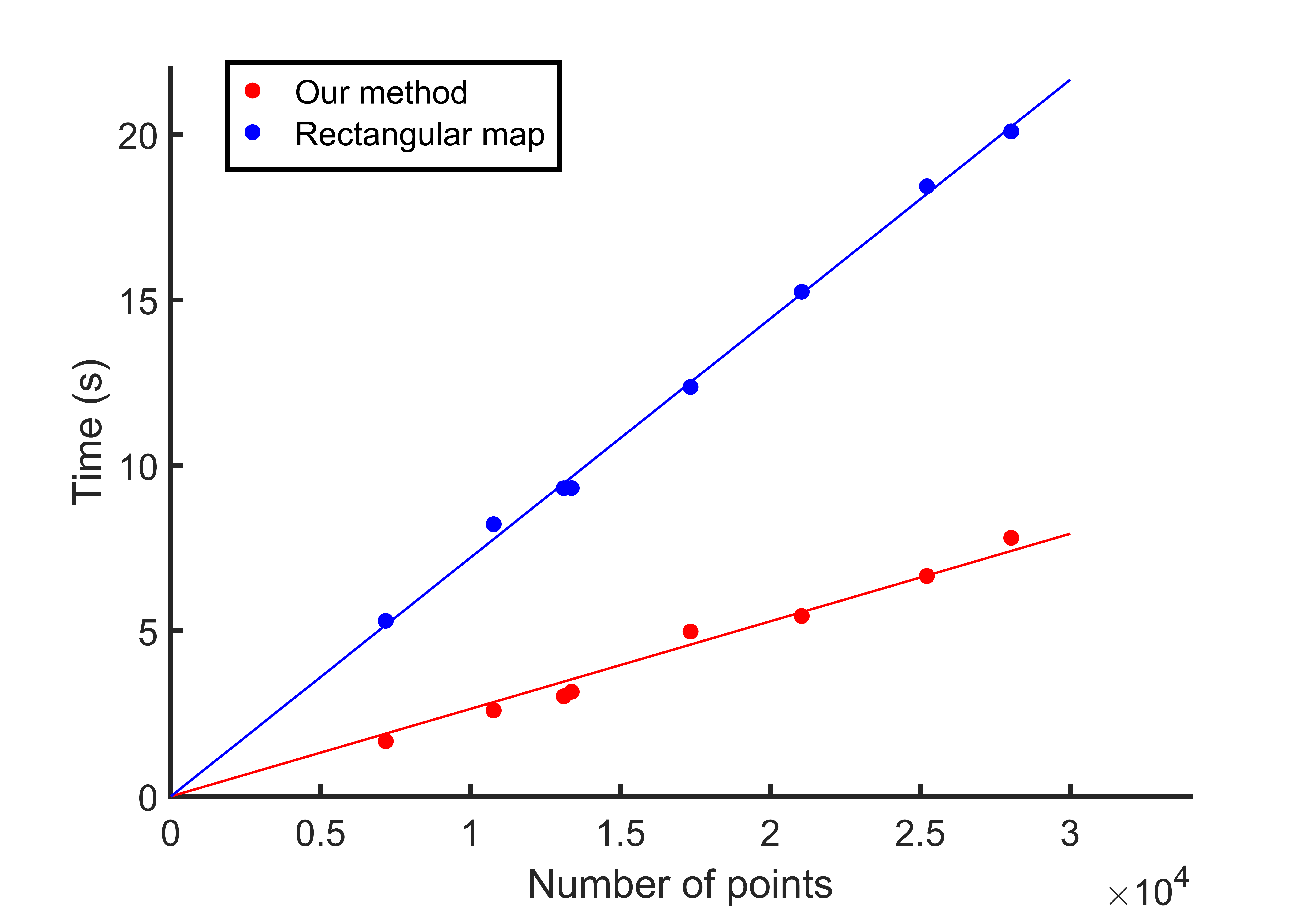}
    \caption{The computation time for the proposed free-boundary conformal parameterization method and the rectangular conformal parameterization method~\cite{meng2016tempo} for point clouds. The markers correspond to the point cloud models in Table~\ref{table:free_boundary}. The best-fit lines for the two methods are also provided.}
    \label{fig:comparison_rect_timing}
\end{figure}

Fig.~\ref{fig:comparison_rect_timing} shows the computation time for our proposed parameterization method and the rectangular parameterization method~\cite{meng2016tempo}. It can be observed that the time required increases approximately linearly with the number of points for both methods. More specifically, the slope for our method is $\approx 2.6\times 10^{-4}$, and that for the rectangular method is $\approx 7.2 \times 10^{-4}$. This suggests that our method is over 60\% faster than the rectangular method on average. The improvement in computation time achieved by our method can be explained by the fact that our method only requires solving the linear system in Eq.~\eqref{eqt:DNCP_PC}, while the rectangular method involves not only solving a linear system to map the point cloud onto a square but also optimizing the height of the rectangular domain to achieve conformality.

Besides, note that the point cloud Laplacian in Eq.~\eqref{eqt:LBoperator_pc} can be used for calculating the Dirichlet energy $E_D(f)$ (Eq.~\eqref{eqt:Cotangentmatrix}) of a point cloud mapping $f:\mathcal{P} \to \mathbb{R}^2$. Similarly, we can replace the cotangent weight in Eq.~\eqref{eqt:Cotangentweights_pc} with the locally authalic weight~\cite{desbrun2002intrinsic} and calculate the locally authalic Chi energy $E_{\chi}(f)$ to measure the local 1-ring area distortion of a point cloud mapping $f$:
\begin{equation}
    E_{\chi}(f) = \sum_{v_i \in \mathcal{P}} \sum_{v_j \in \mathcal{N}^{\mathcal{V}}(v_i)} \frac{\cot \gamma_{ij} + \cot \delta_{ij}}{|v_i - v_j|^2} |f(v_i) - f(v_j)|^2,
\end{equation}
where $\gamma_{ij}$ and $\delta_{ij}$ are the two angles at a vertex $v_j$ in the approximated 1-ring vertex neighborhood $\mathcal{N}^{\mathcal{V}}(v_i)$ of the vertex $v_i$ (see~\cite{desbrun2002intrinsic} for more details). Table~\ref{table:area_distortion} records the value of $E_{\chi}$ for our proposed method and the rectangular parameterization method~\cite{meng2016tempo}. By considering the ratio $E_{\chi}(f_{\text{ours}}) / E_{\chi}(f_{\text{rect}})$, where $f_{\text{ours}}$ and $f_{\text{rect}}$ are our parameterization and the rectangular parameterization respectively, it can be observed that our method reduces the Chi energy by over 30\% on average. This suggests that our proposed method is more advantageous than the rectangular parameterization method in terms of not only the conformality and the computational cost but also the local area distortion.

\begin{table}[t]
    \centering
    \begin{tabular}{c|c|c|c}
        Model & $E_{\chi}(f_{\text{ours}})$ & $E_{\chi}(f_{\text{rect}})$ & $E_{\chi}(f_{\text{ours}}) / E_{\chi}(f_{\text{rect}})$\\ \hline  
        Cloth & 1.2246e+04 & 2.3857e+04 & 0.5133 \\ 
        Julius & 2.6134e+04 & 5.8201e+04 & 0.4490 \\ 
        Niccol\`o da Uzzano & 4.0008e+04 & 5.0443e+04 & 0.7931 \\ 
        Max Planck & 2.4843e+04 &5.7683e+04 & 0.4307 \\ 
        Chinese lion & 3.4569e+04 & 2.7245e+04 & 1.2688 \\  
        Sophie & 4.2953e+04 & 7.9225e+04 & 0.5422  \\ 
        Alex & 9.0598e+04 & 1.0714e+05 & 0.8456 \\ 
        Twisted hemisphere & 2.4911e+04 & 3.6771e+04 & 0.6775 \\ 
    \end{tabular}
    \caption{The local area distortion of the point cloud parameterizations. $E_{\chi}(f_{\text{ours}})$ and $E_{\chi}(f_{\text{rect}})$ are the locally authalic Chi energy for our free-boundary parameterization method and the rectangular parameterization method~\cite{meng2016tempo} respectively.}
    \label{table:area_distortion}
\end{table}

\subsection{Comparison with other point cloud Laplacian approximation schemes}
One may also wonder whether some other existing approximation schemes of the Laplace--Beltrami operator can lead to a more accurate free-boundary conformal parameterization when compared to our proposed approximation scheme in Eq.~\eqref{eqt:LBoperator_pc}. Here we apply Algorithm~\ref{algo:DNCP_PC} with the local mesh method~\cite{lai2013local} and the moving least squares (MLS) method~\cite{liang2012geometric} to obtain free-boundary parameterizations and evaluate the conformality of the results. Note that in the MLS method, the Laplacian matrix is approximated using a linear combination of derivatives obtained from a local parametric approximation of the kNN of each point. At a boundary point $v_i$, the local coordinate system constructed using the kNN $\mathcal{N}_k(v_i)$ only consists of points from one side of $v_i$ and hence the Laplacian approximation is highly inaccurate. Therefore, here we evaluate the performance of the MLS method by constructing a Laplacian with the MLS method used for the interior points and the local mesh method used for the boundary points.

\begin{figure}[t!]
    \centering
    \includegraphics[width=\textwidth]{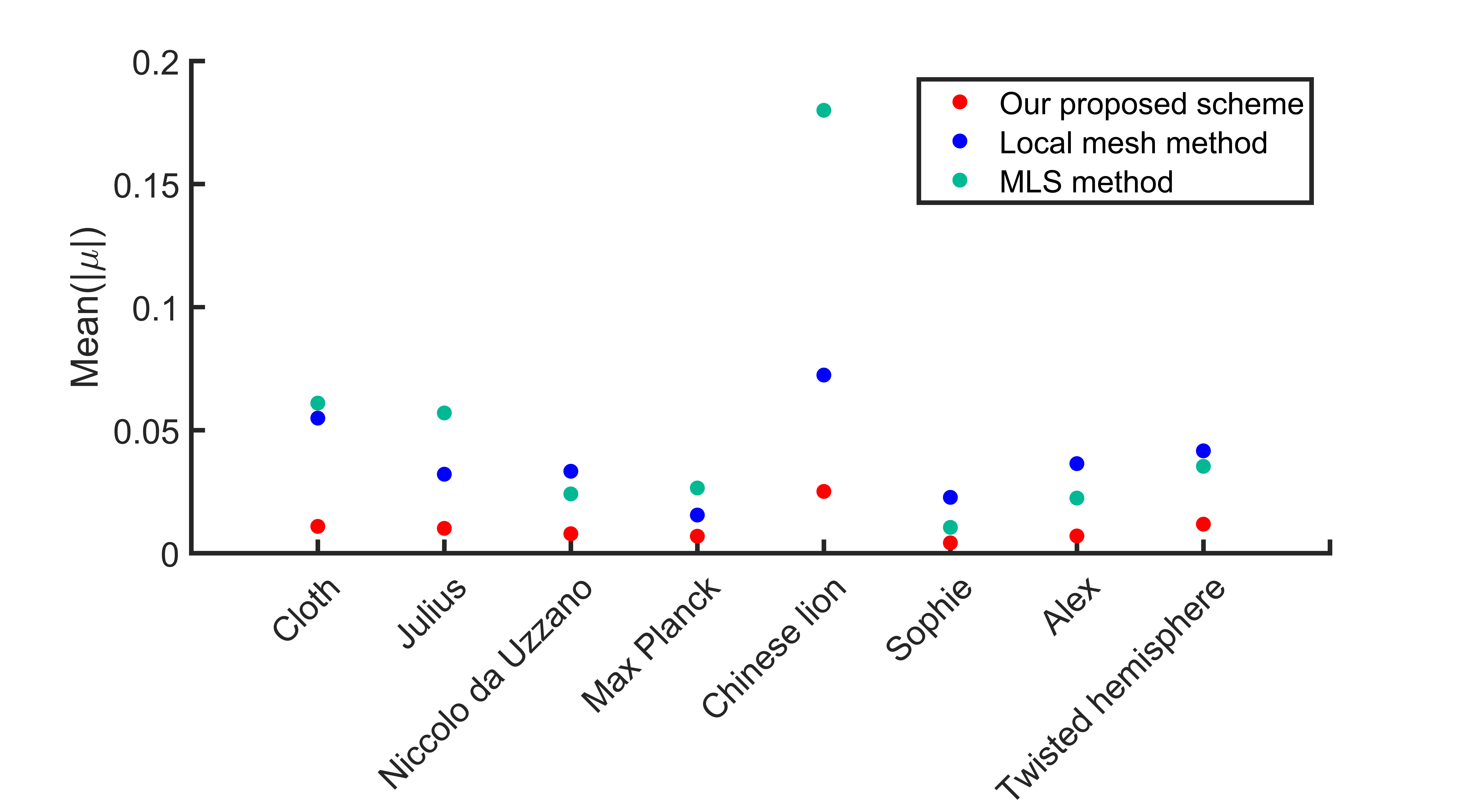}
    \caption{Comparison between our proposed scheme (Eq.~\eqref{eqt:LBoperator_pc}), the local mesh method~\cite{lai2013local} and the moving least squares (MLS) method~\cite{liang2012geometric} for approximating the point cloud Laplacian. For each point cloud model, we apply Algorithm~\ref{algo:DNCP_PC} with the two Laplacian approximation schemes to compute the free-boundary conformal parameterization, and record the mean of the norm of the point cloud Beltrami coefficients (PCBC) $|\mu|$.}
    \label{fig:comparison_laplacian}
\end{figure}

As shown in Fig.~\ref{fig:comparison_laplacian}, the conformal distortion of the parameterizations achieved by Algorithm~\ref{algo:DNCP_PC} with our proposed approximation scheme is lower than that achieved by Algorithm~\ref{algo:DNCP_PC} with the local mesh method and the MLS method. More quantitatively, the conformal distortion achieved by our proposed approximation scheme is 70\% lower than that by both the local mesh method and the MLS method. This suggests that our proposed approximation scheme is important for yielding an accurate free-boundary conformal parameterization of point clouds.

\subsection{Parameterizing noisy point cloud data}
\begin{figure}[t]
    \centering
    \includegraphics[width=\textwidth]{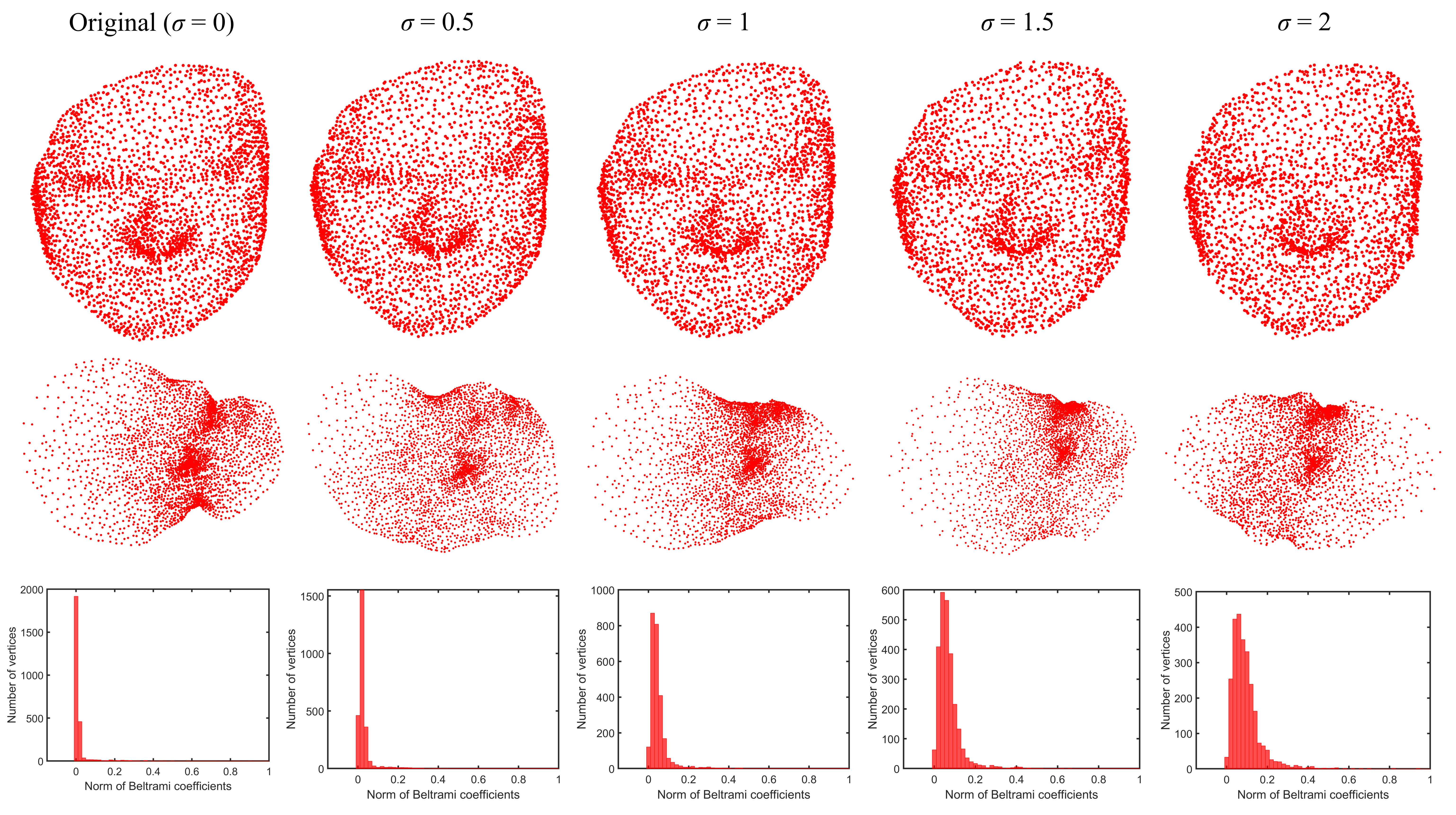}
    \caption{Free-boundary conformal parameterization of noisy point clouds with a Gaussian noise. The leftmost column shows the real-world facial point cloud obtained using the Kinect 3D scanner~\cite{meng2016tempo}, the parameterization result and the histogram of the norm of the point cloud Beltrami coefficient (PCBC) $|\mu|$. The other columns correspond to the noisy point clouds with different $\sigma$.}
    \label{fig:noisy_experiment}
\end{figure}

It is natural to ask whether the proposed parameterization method works well for noisy point clouds. Here, we consider a real-world facial point cloud obtained using the Kinect 3D scanner~\cite{meng2016tempo} (see Fig.~\ref{fig:noisy_experiment}, leftmost). We apply the proposed parameterization method on this point cloud, and the histogram of the norm of the PCBC $|\mu|$ shows that the parameterization is highly conformal. To further study the performance of the proposed parameterization method, we consider adding different level of noise to the facial point cloud. Specifically, we add a Gaussian noise with mean 0 and standard deviation $\sigma = 0.5, 1, 1.5, 2$ using the MATLAB's \texttt{normrnd} function. Fig.~
\ref{fig:noisy_experiment} shows the noisy facial point clouds and the parameterization results. Note that even for the examples with a large $\sigma$, the peaks of the histograms of $|\mu|$ are still close to 0, which suggests that the conformal distortion is satisfactory. 

\begin{figure}[t]
    \centering
    \includegraphics[width=0.9\textwidth]{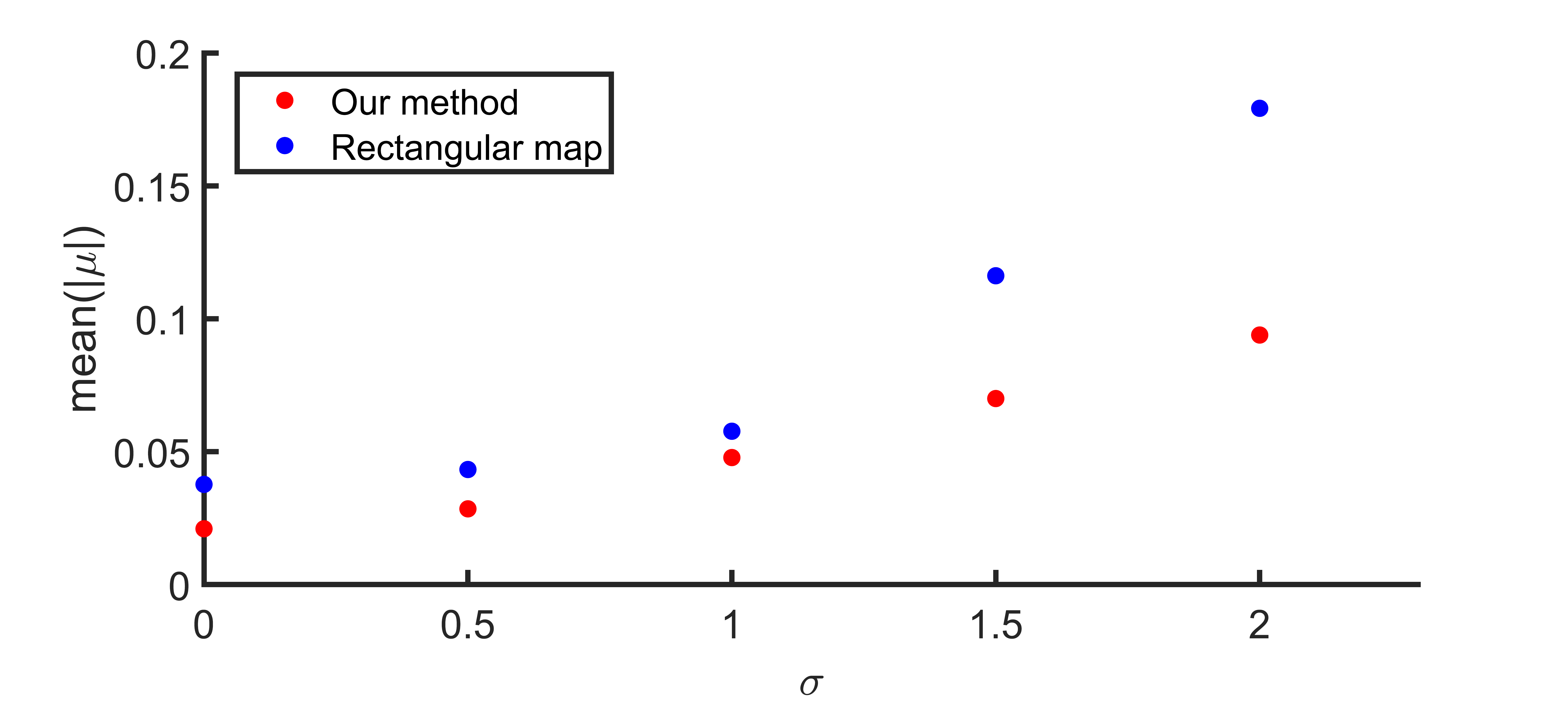}
    \caption{Comparison between our proposed parameterization method and the rectangular parameterization method~\cite{meng2016tempo} for parameterizing noisy point clouds (see Fig.~\ref{fig:noisy_experiment} for the point clouds with different noise parameter $\sigma$). For each example and each method, the mean of the norm of the point cloud Beltrami coefficient (PCBC) $|\mu|$ is recorded.}
    \label{fig:noisy_comparison}
\end{figure}

We further compare the conformal distortion achieved by our proposed method and the rectangular parameterization method~\cite{meng2016tempo} for parameterizing these noisy point clouds. As shown in Fig.~\ref{fig:noisy_comparison}, our proposed parameterization method results in better conformality when compared to the rectangular parameterization method for all levels of noise. This demonstrates the effectiveness of our method for handling noisy point cloud data.

\section{Application to point cloud meshing}\label{sect:application}
\begin{figure}[t!]
    \centering
    \includegraphics[width=0.8\textwidth]{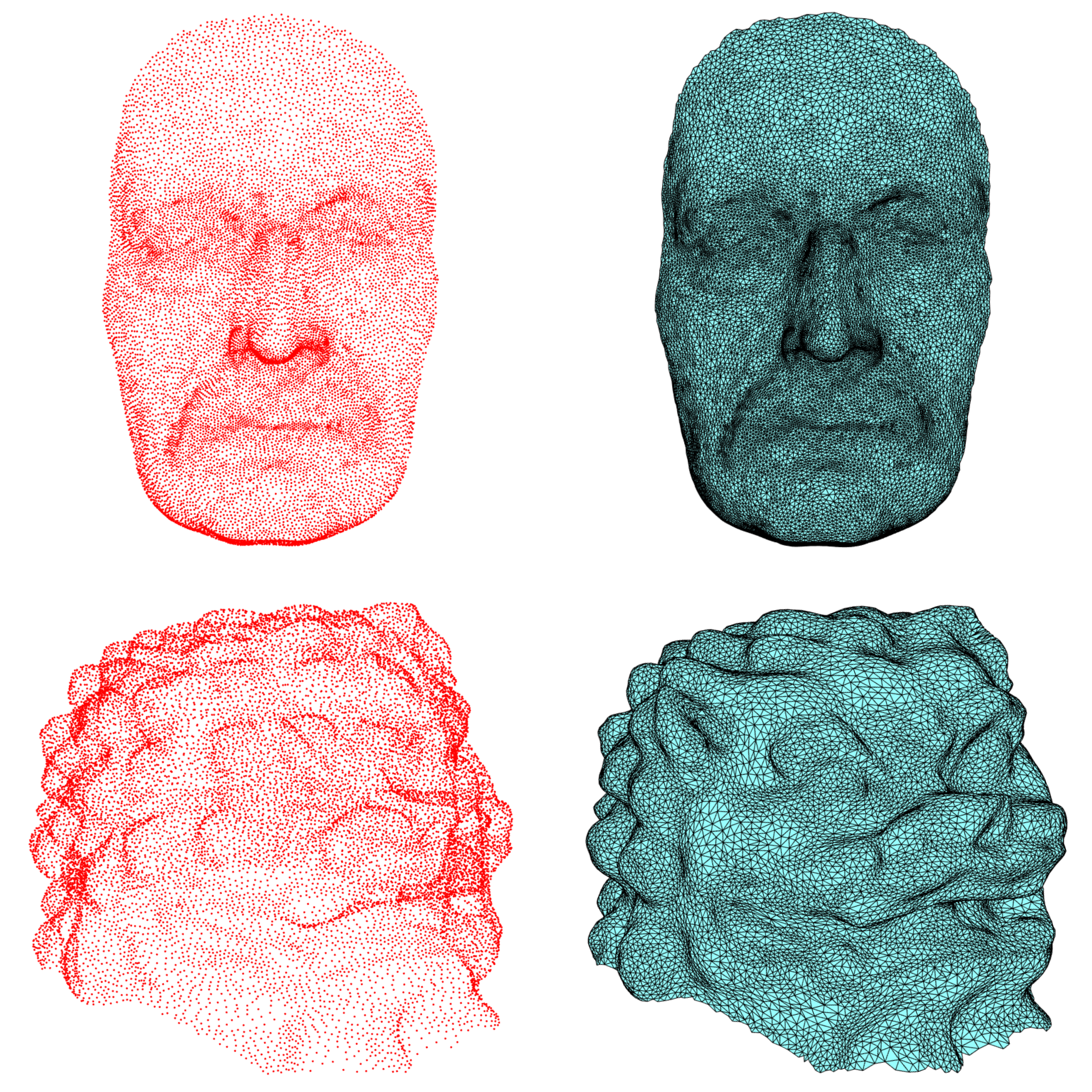}
    \caption{Meshing point clouds via our proposed free-boundary conformal parameterization method. Left: The input point clouds. Right: The resulting triangular meshes.}
    \label{fig:meshing_result}
\end{figure}

The proposed free-boundary conformal parameterization method can be used for meshing disk-type point clouds. More specifically, after parameterizing a point cloud onto the plane, we can compute a 2D Delaunay triangulation of all points, which induces a mesh structure on the input point cloud. Note that every non-boundary edge $\mathbf{e}$ in a 2D Delaunay triangulation is shared by exactly two triangles, in which the two angles $\alpha, \beta$ opposite to $\mathbf{e}$ always satisfy the following property:
\begin{equation}
    \alpha+\beta \leq \pi.
\end{equation}
In other words, angles that are too acute or too obtuse are avoided as much as possible in 2D Delaunay triangulations. Such high-quality triangulations are desirable in many practical applications. As demonstrated by the numerical experiments presented in Section~\ref{sect:experiment}, the proposed free-boundary conformal parameterization method results in a lower conformal distortion when compared to other parameterization approaches. Therefore, the above-mentioned nice property of the 2D Delaunay triangulations is well-preserved in the resulting triangular meshes of the 3D point clouds via our parameterization method. Fig.~\ref{fig:meshing_result} shows two examples of meshing point clouds via our parameterization method, from which it can be observed that the resulting triangles are highly regular. 

For a more quantitative assessment, we consider the \emph{Delaunay ratio} of a point cloud triangulation~\cite{choi2016spherical}:
\begin{equation}
    r = \frac{\text{Number of non-boundary edges satisfying $\alpha+\beta \leq \pi$}}{\text{Total number of non-boundary edges}}.
\end{equation}
Table~\ref{table:delaunay_ratio} records the Delaunay ratio of the point cloud triangulations created via our free-boundary conformal parameterization method and the rectangular conformal parameterization method~\cite{meng2016tempo}. In all experiments, our free-boundary parameterization approach produces triangular meshes with a higher quality when compared to the rectangular boundary parameterization approach. This shows that our proposed free-boundary conformal parameterization method is more advantageous for point cloud meshing.

\begin{table}[t!]
    \centering
    \begin{tabular}{c|c|c}
        Model & Meshing via our method & Meshing via rectangular map~\cite{meng2016tempo} \\ \hline
        Cloth & 0.9991 & 0.9814\\
        Julius & 1 & 0.9987\\
        Niccol\`o da Uzzano & 0.9979 & 0.9935\\
        Max Planck & 0.9984 & 0.9670\\
        Chinese lion & 0.9918 & 0.9805\\ 
        Sophie & 0.9994 & 0.9916\\
        Alex & 0.9980 & 0.9939\\
        Twisted hemisphere & 0.9955 & 0.9882\\
    \end{tabular}
    \caption{The Delaunay ratio $r$ of the point cloud triangulations created via our proposed free-boundary conformal parameterization method and the rectangular conformal parameterization method~\cite{meng2016tempo}.}
    \label{table:delaunay_ratio}
\end{table}

\section{Extending the proposed parameterization method using partial welding} \label{sect:extension}
In our recent work~\cite{choi2020parallelizable}, we proposed a parallelizable algorithm for the conformal parameterization of triangulated surfaces using the idea of partial welding. Here, we show that the partial welding method can be naturally extended to point clouds, thereby providing a more flexible way of computing the free-boundary conformal parameterizations of point clouds.

The partial welding method~\cite{choi2020parallelizable} is outlined below. Let $S_1, S_2$ be two discretized domains on the complex plane with oriented boundary vertices $\partial S_1 =$  $\{a_0,a_1,\dots, a_{n_1}\}$, $\partial S_2 = \{b_0,b_1,\dots, b_{n_2}\}$ and a partial correspondence 
\begin{equation}
    a_i\leftrightarrow b_i, \ \ \ i = 0, 1, \dots, k,
\end{equation}
where $k\leq \min(n_1, n_2)$. The partial welding method finds two conformal maps $\phi_1:S_1 \to \overline{\mathbb{C}}$ and $\phi_2:S_2 \to \overline{\mathbb{C}}$ such that $\phi_1(a_i) = \phi_2(b_i)$ for all $i = 0, 1, \dots, k$. In other words, the two domains are glued conformally along the corresponding vertices. To achieve this, the method first computes a series of conformal maps to map $S_1$ and $S_2$ onto the upper half-plane and the lower half-plane respectively, with $a_0, \dots, a_k$ mapped to the upper half of the imaginary axis (denoted the transformed vertices as $A_0, \dots, A_{n_1}$) and $b_0, \dots, b_k$ mapped to the lower half of the imaginary axis (denoted the transformed vertices as $B_0, \dots, B_{n_2}$). The method then finds another series of conformal maps such that each pair of corresponding vertices $(A_i, B_i)$ are mapped to a point on the real axis, thereby gluing the two domains (denoted the transformed vertices as $\tilde{A}_0, \dots, \tilde{A}_{n_1}$ and $\tilde{B}_0, \dots, \tilde{B}_{n_2}$). In other words, the two desired maps $\phi_1, \phi_2$ are constructed by a composition of these conformal maps. Using this idea of gluing two domains based on a partial correspondence, one can compute a conformal parameterization of a simply-connected open surface by first partitioning it into multiple domains, then conformally mapping each domain onto the complex plane and finally gluing all flattened domains successively. Readers are referred to~\cite{choi2020parallelizable} for more details.

While the partial welding method is developed for the parameterization of triangulated surfaces in~\cite{choi2020parallelizable}, we note that the above-mentioned procedures only involve the boundary points of each domain but not the triangulations. This suggests that the partial welding method is naturally applicable to our point cloud parameterization problem.

\begin{figure}[t]
    \centering
    \includegraphics[width=0.8\textwidth]{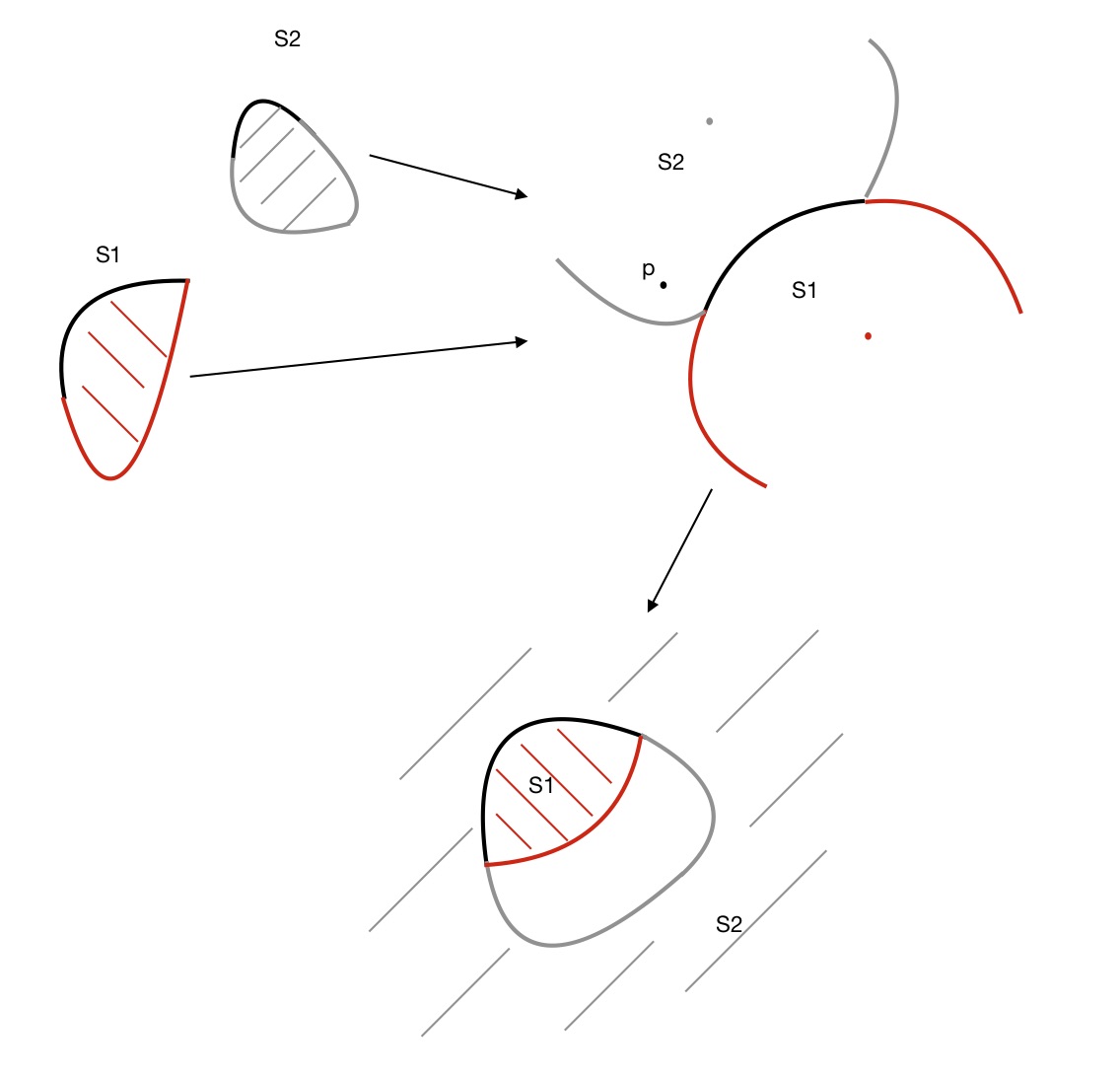}
    \caption{An illustration of the potential occurrence of an unbounded domain under the final M\"obius transformation step in~\cite{choi2020parallelizable}. The two domains $S_1, S_2$ are to be welded partially, with the weld paths highlighted in black. After the welding process, the two domains are welded along the weld paths, with the three points $\left(\tilde{A}_{n_1 +1}, \tilde{B}_{n_2+1},  p = \frac{1}{2}(\tilde{A}_{n_1+2} +\tilde{B}_{n_2+2}) \right)$ marked in different colors. Since $p$ lies inside the transformed $S_2$, a M\"obius transformation sending $p$ to $\infty$ will map the transformed $S_2$ to an unbounded domain.}
    \label{fig:trouble}
\end{figure}

Note that in the original partial welding algorithm~\cite{choi2020parallelizable}, two auxiliary points $a_{n_1+1} = \frac{1}{n_1} \sum_{j=1}^{n_1} a_j$ and $a_{n_1+2} = \infty$ are appended to $\partial S_1$. Similarly, two auxiliary points $b_{n_2+1} = \frac{1}{n_2} \sum_{j=1}^{n_2} b_j$ and $b_{n_2+2} = \infty$ are appended to $\partial S_2$. These auxiliary points are used at the last step of the algorithm in~\cite{choi2020parallelizable}. More specifically, the last step considers a M\"obius transformation that sends $\left(\tilde{A}_{n_1 +1}, \tilde{B}_{n_2+1}, p = \frac{1}{2}(\tilde{A}_{n_1+2} +\tilde{B}_{n_2+2}) \right)$ to $(-1,1,\infty)$ in order to normalize the transformed boundary shapes. However, we notice that $p$ may lie inside the interior of one of the transformed domains in some rare cases. Pushing such $p$ to $\infty$ will map a bounded domain to an unbounded domain (see Fig.~\ref{fig:trouble} for an illustration), which is undesirable. 

To overcome this potential problem, here we propose the following modification of the final M\"obius transformation step. If $p$ lies inside the interior of either the transformed $\partial S_1$ or $\partial S_2$, we consider the boundary $\Gamma = (\partial S_1 \cup \partial S_2 - \text{weld path})\cup (\text{endpoints of weld path}) $ formed by the red and grey curves in the top right panel of Fig.~\ref{fig:trouble}. Note that $\Gamma$ must be a simple closed curve as all the previous maps are conformal. This suggests that we only need to find an interior point $q$ in the bounded Jordan domain $\tilde{\Omega}$ formed by $\Gamma$, and two other points $\Gamma(1),\Gamma(l)\in\Gamma$ where $l\approx |\Gamma|/2\in\mathbb{N}$. To find an interior point in $\tilde{\Omega}$, we simply use a minimal axis-aligned bounding box for $\tilde{\Omega}$, i.e. $[\min (\text{Re}(\Gamma)), \max (\text{Re}(\Gamma))]\times [\min (\text{Im}(\Gamma)), \max (\text{Im}(\Gamma))]$. Then we draw a vertical line $x = \frac{1}{2}(\min (\text{Re}(\Gamma))+\max (\text{Re}(\Gamma)))$ and compute its intersections with $\Gamma$. We sort the intersections by their distance to a point that lies on the line but outside the bounding box. We then take the midpoint of the first two intersections as $q$, which must lie inside $\text{int}(\tilde{\Omega})$. Finally, the final M\"obius transformation in the original partial welding algorithm~\cite{choi2020parallelizable} can be replaced with a M\"obius transformation sending $(\Gamma(1),\Gamma(l), q)$ to $(-1,1,\infty)$. This ensures that the welded boundary shapes can be normalized without being mapped to an unbounded domain. 

\begin{figure}[t]
    \centering
    \includegraphics[width=\textwidth]{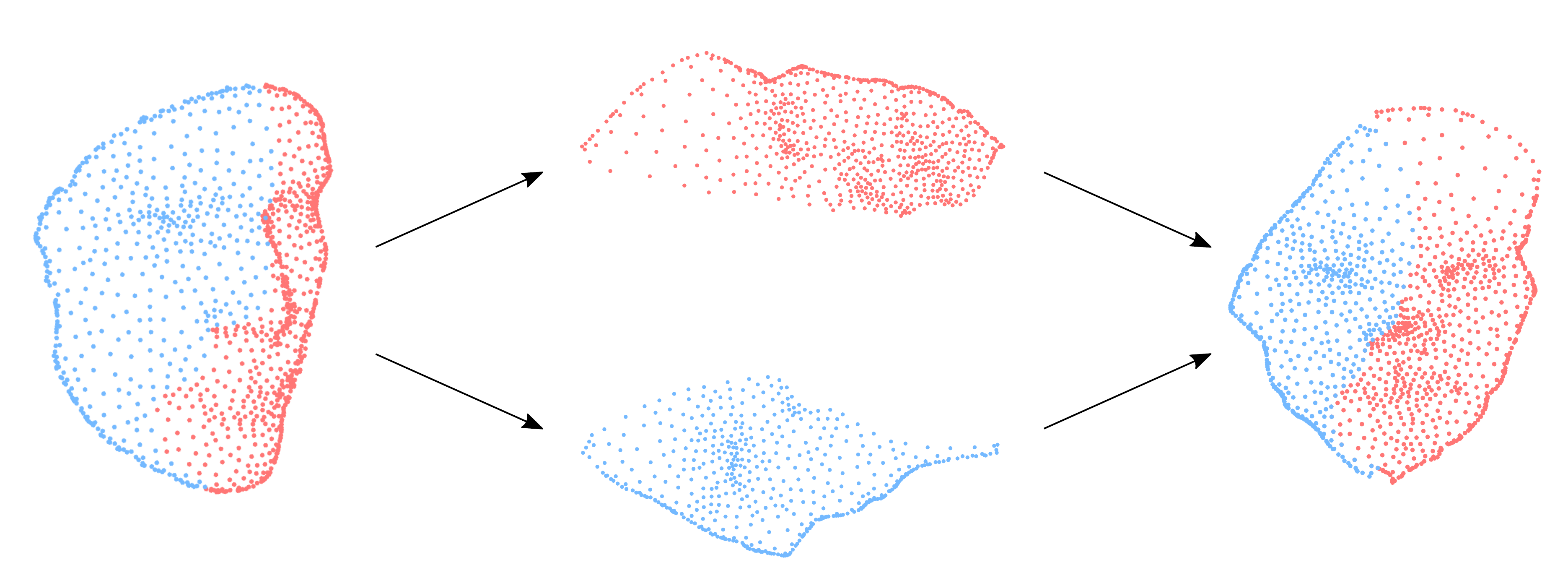}
    \caption{An illustration of the proposed free-boundary conformal parameterization method for point clouds via partial welding. Given a human face point cloud, we first partition it into subdomains. We then conformally flatten each subdomain onto the complex plane using Algorithm~\ref{algo:DNCP_PC}. Finally, we glue all flattened subdomains using partial welding to obtain the global free-boundary conformal parameterization.}
    \label{fig:pc_welding_illustration}
\end{figure}

An illustration of the proposed free-boundary conformal parameterization algorithm for point clouds via partial welding is shown in Fig.~\ref{fig:pc_welding_illustration}. Given a point cloud $\mathcal{P} = \{v_i\}_{i=1}^n$, we first partition it into $m$ subdomains $\mathcal{P}_1, \mathcal{P}_2, \dots, \mathcal{P}_m$. We then compute a conformal parameterization $\varphi_i:\mathcal{P}_i \to \mathbb{C}$ for each subdomain using Algorithm~\ref{algo:DNCP_PC}. We remark that for each point $v^i_{j}$ in $\mathcal{P}_i$, where $j = 1,2,\dots, |\mathcal{P}_i|$, we exclude all points in its $k$-nearest neighbors ${N}^{k,i}_j$ that are not in $\mathcal{P}_i$ for computing $\varphi_i$. Note that the parameterizations $\varphi_1, \dots, \varphi_m$ are independent of each other and hence can be computed in parallel. Once all subdomains are flattened, we can use the partial welding method as described in~\cite{choi2020parallelizable} with the above-mentioned modification of the final M\"obius transformation step to glue all subdomains based on the partial correspondences of their boundaries, thereby yielding a global free-boundary conformal parameterization of $\mathcal{P}$. The proposed free-boundary conformal parameterization method via partial welding is summarized in Algorithm~\ref{algo:PGCP_PC}. 

\begin{algorithm2e}[h!]
\label{algo:PGCP_PC}
\KwIn{A point cloud $\mathcal{P}=\{v_i:1\leq i\leq N\}$ with disk topology with boundary $\partial \mathcal{P}$ and the number of subdomains $m$.} 
\KwOut{A free-boundary conformal parameterization $f:\mathcal{P} \to \mathbb{C}$.}
\BlankLine

Partition $\mathcal{P}$ into $m$ subdomains $\mathcal{P}_1, \mathcal{P}_2, \dots, \mathcal{P}_m$ with weld paths $\gamma_i, i = 1, \dots, m$\;

\For{$i=1:m$}{
    Extract the boundary $\partial \mathcal{P}_i = \gamma_i \cup (\partial \mathcal{P}\cap \mathcal{P}_i)$ for each subdomain\;
    
    \For{$j=1:|\mathcal{P}_i|$}{
    
        \If{$\exists \ v\in {N}^{k,i}_j \setminus \mathcal{P}_i$}{
            Delete $v$ from ${N}^{k,i}_j$\;
            
            Do Line 3--6 in Algorithm \ref{algo:DNCP_PC}\;
        }
        
        Do Line 7--9 in Algorithm \ref{algo:DNCP_PC}\;
    }
    
    Do Line 10--12 in Algorithm~\ref{algo:DNCP_PC} to get a free-boundary conformal parameterization $\varphi_i:\mathcal{P}_i \to \mathbb{C}$. Record the Laplacian matrix and the mapped boundary $\varphi_i(\partial \mathcal{P}_i)$\; 
}

Perform partial welding~\cite{choi2020parallelizable} on the boundaries $\varphi_1(\partial \mathcal{P}_1)$, $\varphi_2(\partial \mathcal{P}_2)$, \dots, $\varphi_m(\partial \mathcal{P}_m)$ obtained in Line 9 with the modified final M\"obius transformation step\;

Use the Laplacian matrices obtained in Line 9 to compute a conformal parameterization $\widetilde{\varphi}_i: \mathcal{P}_i \to \mathbb{C}$ for each $\mathcal{P}_i$ with the welded boundary constraints\;

Combine $\widetilde{\varphi}_1(\mathcal{P}_1), \widetilde{\varphi}_2(\mathcal{P}_2), \dots, \widetilde{\varphi}_m(\mathcal{P}_m)$ to form the final free-boundary conformal parameterization $f:\mathcal{P} \to \mathbb{C}$\;

\caption{Free-boundary conformal parameterization of point clouds via partial welding}
\end{algorithm2e}

\begin{figure}[t]
    \centering
    \includegraphics[width=\textwidth]{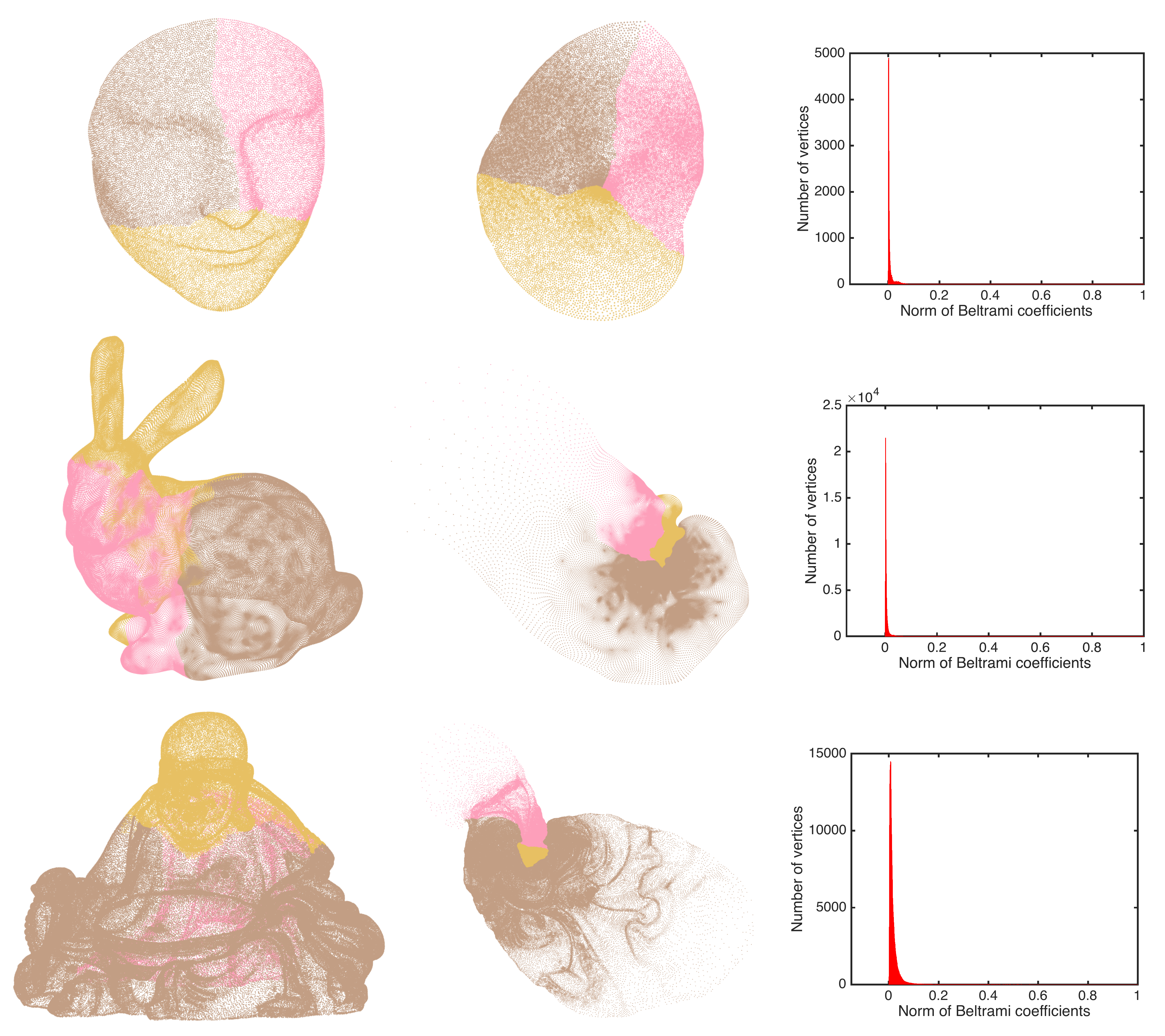}
    \caption{Examples of free-boundary conformal parameterizations of point clouds produced by the proposed method with partial welding (Algorithm~\ref{algo:PGCP_PC}). Left: The input point clouds. Middle: The parameterization results. Right: The histograms of the norm of the point cloud Beltrami coefficients $|\mu|$. The colors indicate the corresponding subdomains in the input point clouds and the parameterization results. For the first two models, the kNN parameter is set to be $k = 25$. As the last model is much denser, the kNN parameter is set to be $k=35$.}
    \label{fig:welding_result}
\end{figure}

We remark that since the computations of all $\varphi_i$ are independent, the input parameters $k, c_1, c_2$ in Algorithm~\ref{algo:DNCP_PC} for computing each $\varphi_i$ can be set differently. In other words, the use of partial welding in Algorithm~\ref{algo:PGCP_PC} allows us to have a more flexible choice of the local approximation parameters for handling regions with different geometry.

Fig.~\ref{fig:welding_result} shows several examples of free-boundary conformal parameterization of point clouds produced by Algorithm~\ref{algo:PGCP_PC}. It can be observed that different subdomains can be handled separately and then glued seamlessly to form the final free-boundary conformal parameterization. The histograms of the norm of the Beltrami coefficients $|\mu|$ show that the parameterizations are highly conformal.

\section{Discussion} \label{sect:discussion}
While mesh parameterization methods have been widely studied over the past several decades, the parameterization of point clouds is much less understood. In this work, we have proposed a method for computing free-boundary conformal parameterizations of point clouds with disk topology. More specifically, we develop a novel approximation scheme of the point cloud Laplacian, which allows us to extend the DNCP mesh parameterization method~\cite{desbrun2002intrinsic} for point clouds with disk topology. The flexibility of the proposed parameterization method can be further enhanced with the aid of partial welding~\cite{choi2020parallelizable}. The proposed method is capable of handling a large variety of point clouds and achieves better conformality when compared to prior point cloud parameterization approaches. The improvement in the conformality makes the proposed point cloud parameterization method suitable for practical applications such as point cloud meshing.

While we have used a universal $k$ in our experiments, one can also use a varying $k$ for constructing the $k$-nearest neighborhood for different points without altering any other steps in the proposed parameterization algorithm. By setting $k$ based on the properties such as curvature and density of the input point cloud, one may be able to further improve the parameterization result. Also, note that as described in~\cite{choi2020parallelizable}, the mesh-based partial welding method for free-boundary conformal parameterization can be modified for achieving other prescribed boundary shapes such as a circle. Analogously, Algorithm~\ref{algo:PGCP_PC} should also be able to be modified for achieving other prescribed boundary shapes in the resulting parameterization, thereby leading to an improvement in conformality and flexibility for such parameterization problems.

As for possible future works, it will be interesting to consider combining the idea of tufted cover~\cite{sharp2020laplacian} with the proposed angle criterion to further improve the approximation of the point cloud Laplacian. We also plan to explore the use of the proposed parameterization method for point cloud registration and shape analysis, and extend the proposed parameterization method for the conformal parameterization of point clouds with some other underlying surface topology. For instance, we should be able to extend Algorithm~\ref{algo:PGCP_PC} for parameterizing multiply-connected point clouds. More specifically, we can partition a multiply-connected point cloud into simply-connected subdomains and compute the conformal parameterization of each of them using the proposed parameterization method, with the partial welding idea utilized for ensuring the consistency between the boundaries of the subdomains. \\

\small{
\noindent \textbf{Acknowledgements} \ This work was supported in part by the National Science Foundation under Grant No.~DMS-2002103 (to Gary P.~T.~Choi), and HKRGC GRF under project ID 2130549 (to Lok Ming Lui).
}

\bibliographystyle{ieeetr}
\bibliography{pcfreebdybib.bib}

\end{document}